\documentstyle[12pt]{article}
\begin{document}\thispagestyle{empty}\noindent\begin{flushright}
OUT--4102--62\\
21 April 1996\\
hep-th/9604128         \end{flushright}\vspace*{2mm}\begin{center}{\Large\bf
On the enumeration of irreducible $k$-fold Euler sums    \\[3pt]
and their roles in knot theory and field theory          \\\vglue 20mm
D.~J.~Broadhurst                                         \\}\vglue 5mm{\tt
D.Broadhurst@open.ac.uk                                  \\}\vglue 5mm{\large
Physics Department, Open University, Milton Keynes MK7 6AA, UK}\\\vglue 5mm{\tt
http://yan.open.ac.uk/                                }\end{center}\vglue 20mm
{\bf Abstract}
A generating function is given for the number, $E(l,k)$, of irreducible
$k$-fold Euler sums, with all possible alternations of sign, and exponents
summing to $l$. Its form is remarkably simple:
$\sum_n E(k+2n,k)\,x^n=\sum_{d|k}\mu(d)\,(1-x^d)^{-k/d}/k$,
where $\mu$ is the M\"obius function. Equivalently, the size
of the search space in which $k$-fold Euler sums of level $l$
are reducible to rational linear combinations of irreducible basis terms
is $S(l,k)=\sum_{n<k}{\lfloor(l+n-1)/2\rfloor\choose n}$.
Analytical methods, using Tony Hearn's REDUCE,
achieve this reduction for the 3698 convergent double Euler sums
with $l\leq44$; numerical methods, using David Bailey's
MPPSLQ, achieve it for the 1457 convergent $k$-fold sums with $l\leq7$;
combined methods yield bases for all remaining search spaces with
$S(l,k)\leq34$.
These findings confirm expectations based on Dirk Kreimer's
connection of knot theory with quantum field theory.
The occurrence in perturbative quantum electrodynamics
of all 12 irreducible Euler sums with $l\leq 7$
is demonstrated.
It is suggested that no further transcendental
occurs in the four-loop contributions to the
electron's magnetic moment.
Irreducible Euler sums are found to occur in explicit analytical
results, for counterterms with up to 13 loops, yielding transcendental
knot-numbers, up to 23 crossings.
\newpage
\newcommand{\df}[2]{\mbox{$\frac{#1}{#2}$}}
\newcommand{\Eq}[1]{~(\ref{#1})}
\newcommand{\Eqq}[2]{~(\ref{#1},\ref{#2})}
\newcommand{\Eqqq}[3]{~(\ref{#1},\ref{#2},\ref{#3})}
\newcommand{\ze}[1]{\,\zeta(#1)}
\newcommand{\zu}[2]{\,U_{#1,#2}}
\newcommand{\zd}[2]{\,\zeta(#1,#2)}
\newcommand{\zt}[2]{\,\zeta(#1,#2,-1)}
\newcommand{\zo}[3]{\,\zeta(#1,-#2,-#3)}
\newcommand{\zn}[3]{\,\zeta(#1,#2,#3)}
\newcommand{\kk}[2]{\,N_{#1,#2}}
\newcommand{\kkk}[3]{\,N_{#1,#2,#3}}
\newcommand{\zs}[1]{\,\zeta^2(#1)}
\newcommand{\zc}[1]{\,\zeta^3(#1)}
\newcommand{\zf}[1]{\,\zeta^4(#1)}
\newcommand{\tm}{\cdot}
\newcommand{\al}[1]{\,\alpha(#1)}
\newcommand{\lp}[1]{\ln^{#1}2}
\newcommand{\ip}[1]{\left\lfloor{#1}\right\rfloor}
\newcommand{\us}[1]{\{1\}_{#1}}
\newcommand{\li}[1]{{\rm Li}_{#1}\!\left(\df12\right)}
\newcommand{\lt}{\ln2}
\newcommand{\ep}{\varepsilon}
\newcommand{\nl}{\nonumber\\&&{}}
\newcommand{\nq}{\nonumber\\}
\newcommand{\cis}[2]{\label{C3#2}\\&\approx&#1\nonumber}
\newcommand{\se}[1]{\sigma_3^{#1}}
\newcommand{\st}[1]{\sigma_2^{#1}}
\newcommand{\so}[1]{\sigma_1^{#1}}
\newcommand{\fk}[4]{\st{#1}\so{#2}\st{#3}\so{#4}}
\newcommand{\sk}[6]{\so{#1}\st{#2}\so{#3}\se{#4}\st{#5}\se{#6}}
\newcommand{\eltft}{\sk22{}{}32}
\newcommand{\vtx}{\circle*{7}}
\newcommand{\ovl}[1]{\overline{#1}}
\newcommand{\bt}[3]{(#1,\ovl{#2},\ovl{#3})}
\newcommand{\bq}[4]{(\ovl{#1},\ovl{#2},\ovl{#3},\ovl{#4})}
\section{Introduction}

Recent progress in number theory~\cite{BBG,BG} interacts strongly
with the connection between knot theory and quantum field theory,
discovered by
Dirk Kreimer~\cite{DK1,DK2},
and intensively investigated to 7 loops~\cite{AI95}
by analytical and numerical techniques. The sequence
of irreducible non-alternating double Euler sums studied in~\cite{BBG}
starts with a level-8 sum
that occurs in the 6-loop renormalization of quantum field
theory~\cite{AI95},
where its appearance is related~\cite{DK2} to the uniquely positive
8-crossing knot $8_{19}$; the sequence of
irreducible non-alternating triple Euler sums
in~\cite{BG} starts with a level-11 sum that occurs at 7 loops,
where its appearance is associated with the uniquely positive hyperbolic
11-crossing knot~\cite{AI95}.

These exciting connections, between number theory, knot theory, and quantum
field theory, led to work with Bob Delbourgo and
Dirk Kreimer~\cite{BDK},
on patterns~\cite{BK} of transcendentals in perturbative expansions
of field theories with local gauge invariance,
and with John Gracey and Dirk Kreimer~\cite{BGK},
on transcendentals generated by all-order~\cite{VPH}
results in field theory,
obtainable in the limit of a large number, $N$,
of interacting fields~\cite{JAG,lnf}.

In the course of the large-$N$ analysis~\cite{BGK},
the number theory in~\cite{BBG} appeared to constitute a severe
obstacle to the development of the connection between knot theory and
field theory. {From} the skeining of link diagrams that encode the flow of
momenta in Feynman diagrams, we repeatedly obtained a family of knots,
associated with the occurrence of irreducible Euler sums
in counterterms.
The obstacle was created by the (indubitably correct)
`rule of 3' discovered in~\cite{BBG}, for
non-alternating sums $\zeta(a,b)=\sum_{n>m}1/n^a m^b$
of level $l=a+b$. The analysis of~\cite{BBG} shows
that non-alternating sums of odd levels are reducible,
while at even level $l=2p+2$ there are $\ip{p/3}$ irreducibles,
where $\ip{\ldots}$ is the integer part.
At levels 8 and 10 this made us very happy, since we had
the 8-crossing knot $8_{19}$ to associate with $\zeta(5,3)$,
and the 10-crossing knot $10_{124}$ to associate with $\zeta(7,3)$.
Thereafter the knots increase in number by a `rule of
two', giving $\ip{p/2}$ knots with $l=2p+4$ crossings.
So there are two 12-crossing knots,
while~\cite{BBG} has only {\em one\/} level-12 irreducible,
and two 14-crossing knots, which {\em is\/}~\cite{BBG}
the number of level-14 irreducibles.

Faced with a 12-crossing knot in search of a number,
we saw two ways to turn: to study 4-fold non-alternating
sums, or 2-fold sums with alternating signs.
The first route is numerically intensive: it soon emerges
that well over 100 significant figures are needed to find
integer relations between 4-fold sums at level 12.
The second route
is analytically challenging; it soon emerges that
at all even levels $l\geq6$ there are relations between alternating double
sums that cannot be derived from any of the identities
given in~\cite{BBG}.

Remarkably, these two routes lead, eventually,
to the {\em same\/} answer.
The extra 12-crossing knot is indeed associated with
the existence of a 4-fold
non-alternating sum, $\zeta(4,4,2,2)=\sum_{n>m>p>q}1/n^4m^4p^2q^2$,
which {\em cannot\/} be reduced to non-alternating sums
of lower levels. It is, equivalently, associated
with the existence of an irreducible
{\em alternating\/} double sum,
$U_{9,3}=\sum_{n>m}\left\{(-1)^n/n^9\right\}\left\{(-1)^m/m^3\right\}$.
The equivalence stems from the unsuspected circumstance that the combination
$\zeta(4,4,2,2)-(8/3)^3\zu93$, and only this combination,
is reducible to non-alternating double sums.
Moreover, $l=12$ is the lowest level at which the reduction
of non-alternating 4-fold sums necessarily entails an alternating
double sum. The `problem pair' of knots are a problem no more.
Their entries in the knot-to-number dictionary~\cite{DK}
record that they led to a new discovery in number theory:
{\em the reduction of non-alternating sums necessarily entails
alternating sums}.

This discovery led me to study the {\em whole\/}
universe of $k$-fold Euler sums, with all possible alternations
of sign, at all levels $l$.
As will be shown in this paper, it is governed by beautifully simple rules,
which might have remained hidden, were it not for Dirk Kreimer's
persistent transformation of the Feynman diagrams of field theory to
produce a {\em pair} of 12-crossing knots.

The remainder of the paper is organized as follows. Section~2
states\footnote{It would be very difficult to {\em prove}.
One cannot even prove that $\zeta^2(3)/\zeta(6)$ is irrational.}
the formula for the number, $E(l,k)$, of irreducible $k$-fold sums,
with all possible alternations of sign, at level $l$. Section~3 outlines
the process by which it was discovered. The anterior numerics of Section~4
describe the high-precision evaluation methods
and integer-relation searches that helped to produce the formula;
the posterior analytics of Section~5 describe computer-algebra
proofs of rigorous upper bounds on $E(l,k)$ that are respected (and
often saturated) by it.
Section~6 summarizes numerical and analytical findings by
listing an instructive choice of concrete bases.
Section~7 shows that all 12 of the irreducible sums with $l\leq7$
appear in perturbative quantum electrodynamics. Section~8 considers
the import, for quantum field theory, for knot theory, and for
number theory, of results obtained by calculations up to level 23,
corresponding to knots with up to 23 crossings,
and to Feynman diagrams with up to 13 loops.

\section{Result}

To specify an alternating $k$-fold Euler sum, one may give
a string of $k$ signs and a string
of $k$ positive integers. It is very\footnote{Without such a
convention, results such as\Eqq{ll}{perm}
become almost unreadable.}
convenient to combine these strings, by defining
\begin{equation}
\zeta(a_1,\ldots,a_k)=\sum_{n_i>n_{i+1}}\prod_{i=1}^k
\frac{({\rm sign}\,a_i)^{n_i}}{n_i^{|\,a_i|}}\,,
\label{zdef}
\end{equation}
for $k>1$,
on the {\em strict\/} understanding that the arguments are non-zero integers,
and that $a_1\neq1$, to prevent a divergence. Hence one
avoids a proliferation of disparate
symbols for the $2^k$ types of $k$\/-fold sum.
The correspondence with the double-sum notations of~\cite{BBG} is
\begin{eqnarray}
\zeta(s,t)&=&\sigma_{\rm h}(t,s)\,,\label{zpp}\\
\zeta(-s,t)&=&\alpha_{\rm h}(t,s)\,,\label{zmp}\\
\zeta(s,-t)&=&-\sigma_{\rm a}(t,s)\,,\label{zpm}\\
\zu{s}{t}\equiv\zeta(-s,-t)&=&-\alpha_{\rm a}(t,s)\,,\label{zmm}
\end{eqnarray}
for {\em positive\/} integers $s$ and $t$, with emphatically {\em no\/}
implication of analytic continuation.

In perturbative quantum field theory, three-loop radiative
corrections~\cite{AFMT,LR} involve
\begin{equation}
\zu31=
\sum_{n>m>0}\frac{(-1)^{n+m}}{n^3m}
=\df12\ze4-2\left\{\li4
+\df{1}{24}\lp2\left(\lp2-\pi^2\right)\right\}\,.\label{U31}
\end{equation}
At six~\cite{5LB,exp,zs6} and seven~\cite{AI95} loops,
counterterms associated with the
$(4,3)$ and $(5,3)$ torus knots, $8_{19}$~\cite{DK2} and
$10_{124}~\cite{AI95}$,
involve $U_{5,3}$ and $U_{7,3}$, whose irreducibility
is equivalent to that of $\zeta(5,3)$ and $\zeta(7,3)$,
respectively.
In higher counterterms, the {\em independent} irreducibility
of $U_{9,3}$ and $U_{7,5}$ is associated~\cite{BGK}
with a {\em pair} of 12-crossing knots.

In what follows, the number of summations, $k$, is referred to
as the {\em depth\/} of the sum\Eq{zdef},
and $l=\sum_i|\,a_i|$  is referred to as its {\em level}.
The set $S_{l,k}$ of convergent sums of level $l$ and depth $k$ has
\begin{equation}
N(l,k)=2^k{l-1\choose k-1}-2^{k-1}{l-2 \choose k-2}
\label{Nlk}
\end{equation}
elements, as is easily proven by induction.
The number of convergent sums at level $l$ is
\begin{equation}
N(l)=\sum_{k=1}^l N(l,k)=4\times3^{l-2}\,,
\label{Nl}
\end{equation}
provided that $l>1$.

{\bf Table~1}:
Euler's triangle\footnote{It seems appropriate to call it {\em Euler's\/}
triangle; see Section 8.3 for the connection with {\em Pascal's}.}
of irreducibles, at level $l$ and depth $k$,
for $l+k\leq32$\,.
\[\begin{array}{l|cccccccccccccccc}
l\backslash k&\,1\,&\,2\,&3&4&5&6&7&8&9&10&11&12&13&14&15&\ldots\\[3pt]\hline
1&1\\
2&1\\
3&1\\
4&&1\\
5&1&&1\\
6&&1&&1\\
7&1&&2&&1\\
8&&2&&2&&1\\
9&1&&3&&3&&1\\
10&&2&&5&&3&&1\\
11&1&&5&&7&&4&&1\\
12&&3&&8&&9&&4&&1\\
13&1&&7&&14&&12&&5&&1\\
14&&3&&14&&20&&15&&5&&1\\
15&1&&9&&25&&30&&18&&6&&1\\
16&&4&&20&&42&&40&&22&&6&&1\\
17&1&&12&&42&&66&&55&&26&&7&&1\\
18&&4&&30&&75&&99&&70&&30&&7&&\ldots\\
19&1&&15&&66&&132&&143&&91&&35&&\ldots\\
20&&5&&40&&132&&212&&200&&112&&\ldots\\
21&1&&18&&99&&245&&333&&273&&\ldots\\
22&&5&&55&&212&&429&&497&&\ldots\\
23&1&&22&&143&&429&&715&&\ldots\\
24&&6&&70&&333&&800&&\ldots\\
25&1&&26&&200&&715&&\ldots\\
26&&6&&91&&497&&\ldots\\
27&1&&30&&273&&\ldots\\
28&&7&&112&&\ldots\\
29&1&&35&&\ldots\\
30&&7&&\ldots\\
31&1&&\ldots
\end{array}\]

The main purpose of this work is to determine,
as in Table~1, the number, $E(l,k)$, of
irreducible sums in $S_{l,k}$, i.e.\ the minimum
number of sums which, together with sums of lesser depth and products
of sums of lower level, furnish a basis for expressing
the elements of $S_{l,k}$ as linear combinations of terms, with
{\em rational\/} coefficients.
The conclusion is that
\begin{equation}
E(l,k)=\delta_{l,2}\delta_{k,1}+\frac{2}{l+k}\sum_{2d|l\pm k}
\mu(d){\frac{l+k}{2d}\choose\frac{l-k}{2d}}\,,
\label{Elk}
\end{equation}
where the summation is over the positive integers $d$
such that $(l\pm k)/2d$ are integers, and is weighted by the
M\"obius function, $\mu(d)$, which vanishes if $d$ is divisible
by the square of a prime and otherwise is $\pm1$, according as
whether $d$ has an even or odd number of prime divisors.
With the exception
of $\ln2$, from $S_{1,1}$, and $\pi^2$, from $S_{2,1}$,
the irreducibles come from $S_{k+2j,k}$, with $j>0$.
Moreover, $S_{k+2j,k}$ contains
the same number of irreducibles as $S_{j+2k,j}$, as illustrated in
Table~1.

{From}\Eq{Elk} one obtains the number of irreducibles at level $l$:
\begin{equation}
E(l)=\sum_{k=1}^l E(l,k)
=\frac{1}{l}\sum_{d|l}\mu(l/d)\left\{F_{d+1}+F_{d-1}\right\}\,,
\label{El}
\end{equation}
in terms of the Fibonacci numbers.
The integer sequence\Eq{El} is tabulated as M0317 by Sloane and
Plouffe~\cite{SP}, who record its origin in the study~\cite{BSD} of
congruence identities.
It shows that the irreducibles are an
exponentially decreasing fraction of the
number of convergent sums at level $l$:
\begin{equation}
\frac{E(l)}{N(l)}\sim\frac{9}{4l}\exp(-\beta l)\,;\quad
\beta=\ln\frac{6}{\sqrt5+1}
\approx0.6174\,.
\label{BtoN}
\end{equation}
This relative sparsity is illustrated in Table~2.

{\bf Table~2}:
The numbers, $E(l)$ and $N(l)$, of irreducibles and sums,
at level $l$, for $l\leq13$\,.
\[\begin{array}{r|rrrrrrrrrrrrr}
l\phantom)&1&2&3&4&5&6&7&8&9&10&11&12&13\\[3pt]\hline
E(l)&1&1&1&1&2&2&4&5&8&11&18&25&40\\
N(l)&1&4&12&36&108&324&972&2916&8748&26244&78732&236196&708588
\end{array}\]

{From}\Eq{Elk} one may obtain the size, $S(l,k)$, of the search space
for sums of level $l$ and depth $k$,
i.e.\ the minimum number of terms that allow one to express
every element of $S_{l,k}$ as a
linear combination, with rational
coefficients. This basis consists of the irreducibles
with level $l$ and depths no greater than $k$,
together with all the independent terms
that are formed from products of sums whose
levels sum to $l$ and whose depths sum to no more than $k$.

One may generate $S(l,k)$ from $E(l,k)$, by expanding
\begin{equation}
\sum_{n=1}^{l_{\rm max}}\left[\sum_{l=1}
^{l_{\rm max}}\left((X_{2,1,1}\,x^2)^l y
+\sum_{k=1}^l\sum_{i=1}^{E(l,k)}X_{l,k,i}\,x^l y^k\right)\right]^n\label{gen}
\end{equation}
to order $l_{\rm max}$ in both $x$ and $y$, where $l_{\rm max}$
is greatest level
required, and $X_{l,k,i}$ serves as a symbol for the $i$th
irreducible in $S_{l,k}$, so that $(X_{2,1,1})^n$ stands for
any rational multiple of $\pi^{2n}$, and hence for $\zeta(2n)$.
Then $S(l,k)$ is obtained by selecting the terms of order $x^l$,
dropping powers of $y$ higher than $y^k$, setting $y=1$, and counting
the length of the resulting expression. This procedure
is easily implemented in REDUCE, Maple, Mathematica, etc.

{\bf Table~3}:
The size, $S(l,k)$, of the search space for sums in $S_{l,k}$,
for $k\leq l\leq15$\,.
\[\begin{array}{l|rrrrrrrrrrrrrrr}
l\backslash k&1&2&3&4&5&6&7&8&9&10&11&12&13&14&15\\[3pt]\hline
1&1\\
2&1&2\\
3&1&2&3\\
4&1&3&4&5\\
5&1&3&6&7&8\\
6&1&4&7&11&12&13\\
7&1&4&10&14&19&20&21\\
8&1&5&11&21&26&32&33&34\\
9&1&5&15&25&40&46&53&54&55\\
10&1&6&16&36&51&72&79&87&88&89\\
11&1&6&21&41&76&97&125&133&142&143&144\\
12&1&7&22&57&92&148&176&212&221&231&232&233\\
13&1&7&28&63&133&189&273&309&354&364&375&376&377\\
14&1&8&29&85&155&281&365&485&530&585&596&608&609&610\\
15&1&8&36&92&218&344&554&674&839&894&960&972&985&986&987
\end{array}\]

Inspection of Table~3 reveals that the sizes satisfy the recurrence relation
\begin{equation}
S(l+1,k)=S(l,k-1)+S(l-1,k)\,,\quad\mbox{for }l>k>1\,.
\label{srec}
\end{equation}
Moreover, the Fibonacci numbers appear at maximum depth:
\begin{equation}
S(l,l)=S(l,l-1)+1=F_{l+1}\,.\label{sfib}
\end{equation}
{From} the Lucas relation between Fibonacci and binomial numbers, one obtains
\begin{equation}
S(l,k)=\sum_{n=0}^{k-1}
{\ip{\frac{l+n-1}{2}}\choose n}\,,\label{Slk}
\end{equation}
as the solution to\Eqq{srec}{sfib}.

Since the process of generating Table~3 from Table~1 is reversible,
formula\Eq{Elk} may be replaced by the simple
statement that at level $l$
the size of the search space increases by the binomial coefficient
\begin{equation}
S(l,k+1)-S(l,k)=
{\ip{\frac{l+k-1}{2}}\choose k}\,,\label{Plk}
\end{equation}
when the depth is increased from $k$ to $k+1$. This pragmatic
formulation\footnote{Computation leaves
no doubt as to the equivalence of\Eq{Elk} and\Eq{Plk},
though it is not yet proven.}
is rather helpful, when using an integer-relation search algorithm,
such as PSLQ~\cite{PSLQ}. An even simpler, though informal,
restatement appears in Section~8.3.

\section{Discovery}

Some brief historical remarks seem in order at this stage, since
my route to\Eq{Elk} in fact {\em began\/} with the observation of 7
members of the Fibonacci sequence\Eq{sfib},
during the course of 800-significant-figure integer-relation
searches entailed by the relation between knot theory and field theory.

Using David Bailey's magnificent MPPSLQ~\cite{MPPSLQ}
implementation
of PSLQ~\cite{PSLQ}, I succeeded in reducing all 1457 sums in
$\{S_{l,k}\mid1\leq k\leq l\leq7\}$ to just 12 numbers and their products.
These irreducible numbers may
conveniently be taken as
\begin{equation}
\ln2,\,\pi^2,\,\{\ze{l}\mid l=3,5,7\},\,\{\al{l}\mid l=4,5,6,7\},
\zu51,\zt51,\zt33,
\label{to7}
\end{equation}
with the polylogarithms
\begin{equation}
\al{l}=\frac{(-\ln2)^l}{l\,!}
\left\{1-\frac{l(l-1)}{12}\left(\frac{\pi}{\ln2}\right)^2\right\}
+\sum_{n=1}^\infty\frac{1}{2^n n^l}
\label{alpha}
\end{equation}
populating the deepest diagonal of Table~1.
As in\Eq{U31}, the definition\Eq{alpha}
postpones the appearance of $\pi^2(\ln2)^{l-2}$ and $(\ln 2)^l$
to depths $l-1$ and $l$, respectively, which is required by\Eq{Plk}.
Moreover $\al1=\al2=0$, and $\al3=\frac78\ze3$, which
is not a new irreducible.

After noting the Fibonacci sequence for the maximum sizes $S(l,l)$,
with $l\leq7$, I sought a combinatoric form for $S(l,k)$.
Formula\Eq{Slk} suggested itself on the empirical basis of the 28
cases with  $1\leq k\leq l\leq7$, and was then submitted
to intense numerical and analytical
tests at higher levels, as indicated in Sections~4 and~5.
I implemented the generator\Eq{gen}, using
the weight and length commands of REDUCE~\cite{RED}, and constructed
Table~1, working backwards from\Eq{Plk}.
Next came the observation that the generating functions for the
$k=2$ and $k=3$ columns of Table~1 have comparable forms:
\begin{equation}
G_2(x)=\df12\left\{1/(1-x)^2-1/(1-x^2)\right\}\,,\quad
G_3(x)=\df13\left\{1/(1-x)^3-1/(1-x^3)\right\}\,.\label{G23}
\end{equation}
The symmetry of Table~1 was vital to the discovery of the simple formula
\begin{equation}
G_k(x)=\sum_{n=0}^\infty E(k+2n,k)\,x^n=
\frac{1}{k}\sum_{d|k}\mu(d)\,(1-x^d)^{-k/d}\,,
\label{Gk}
\end{equation}
which produces formula\Eq{Elk}.
Computation of $\sum_k E(l,k)$, for $3\leq l\leq100$,
revealed that it produced the integers nearest to
$\sum_{d|l}\mu(l/d)\phi^{d}/l$, where $\phi=\frac12(\sqrt5+1)$
is the golden section. The
equivalent form\Eq{El} was obtained
by submitting the integer sequence $E(3),\ldots,E(23)$
to Neil Sloane's helpful on-line~\cite{NJAS} version of~\cite{SP}.
It was noteworthy that this lookup returned the values $E(1)=E(2)=1$,
from the Fibonacci form\Eq{El}, agreeing with the appearance
of $\ln2$, at $l=1$, and $\pi^2$, at $l=2$. Their inclusion in the $k=1$
column of Table~1, above Euler's triangle,
thus became appropriate, as they appear to usher
in the higher transcendentals, in much the same way that
$F_1=F_2=1$ seed the exponential growth of MPPSLQ's CPUtime
(or Fibonacci's rabbits) along the deepest diagonal of Table~3.

It remains to describe yet more probing
numerical and analytical tests, in further support of
the claim in\Eq{Plk}, and its equivalent version in\Eq{Gk}.
Nonetheless, it is hoped that the reader already shares some of my feeling
that these two formul{\ae} are simply too beautiful to be wrong.

\section{Anterior numerics}

\subsection{Numerical evaluation}

Suppose that one wishes to obtain a high-precision approximation to
$S_\infty=\lim_{N\to\infty}S(N)$,
where the truncated sum $S(N)=\sum_{n\leq N}R(n)$
has a summand $R(n)$ with an asymptotic
series in $1/n$, starting at $1/n^{C+1}$, with $C>0$.
{From} $\{S(n)\mid N-M\leq n\leq N\}$, one may form a table
$\{T(n,m)\mid N-M+m\leq n\leq N,\ 0\leq m\leq M\}$, by the procedure
\begin{equation}
T(n,m+1)=\frac{(C+m+n)\,T(n,m)-n\,T(n-1,m)}{C+m}\,,\label{it}
\end{equation}
with $T(n,0)=S(n)$.
The method exploits
the vanishing of $C+m+n-n/(1-1/n)^{C+m}$ as $n\to\infty$.
It takes $M(M+1)/2$ applications of\Eq{it} to produce $T(N,M)$.
Provided that $N>>M\geq1$, and that rounding errors
have been controlled, one obtains the approximation
$S_\infty=T(N,M)+{\rm O}(M!/N^{M+C})$, where the factorial becomes
significant for $N>>M>>C\geq1$.

This elementary and economical method of accelerated convergence
is applicable to every Euler sum of the form\Eq{zdef} that has no argument
equal to unity, in which case repeated appeal to the
Euler-Maclaurin formula~\cite{EM} underwrites
the absence of logarithms in the expansion of the truncation error,
and the outermost summation is of the form
$\sum_n\left\{R(2n)\pm R(2n+1)\right\}$,
with $R(n)={\rm O}(1/n^{a})$, for an argument $a_1=\pm a$.
Thus one should store truncations
after even increments of $n_1$, and
set $C=a_1-1$ for $a_1\geq2$, or $C=|\,a_1|$ for $a_1\leq-1$, in
procedure\Eq{it}.
To obtain the starting values, one has merely to set up a {\em single\/}
loop that updates and stores each layer of the nest
as its particular summation variable, $n_i$, assumes the
even and odd values within the loop.
Thus the evaluation time for a truncation at $N$ of a $k$-fold sum is roughly
proportional to $k N$.

For sums with no unit arguments, one needs therefore only a few lines of
conventional FORTRAN, which may be handed over
to David Bailey's TRANSMP~\cite{TMP} utility, to produce code
that calls his MPFUN~\cite{MPF} multiple-precision subroutines.
As a rule of thumb, the working accuracy should be somewhat better
than the {\em square\/} of the desired output accuracy, when using\Eq{it}.
When, and {\em only\/} when,
rounding errors are so controlled, an output accuracy of {\em very\/}
roughly $M!/N^M$ is achieved by $M$ iterations of\Eq{it},
with input data obtained from looping over $N$ pairs of successive even and
odd integers. For a 4-fold sum,
the accumulation of data takes roughly $50N$ calls of MPFUN subroutines,
and the acceleration of the convergence takes roughly $8M^2$ calls.
Thus, to achieve $P$ significant figures for a 4-fold sum,
one should choose a value of $M$ that keeps the time factor,
$T\approx 50(10^P M!)^{1/M}+8M^2$, close to its absolute minimum.
I find that between 780 and 800 significant figures are reliably
and efficiently achieved with $M=440$ iterations, and truncation
at $N=10^4$, which entails $T\approx2\times10^6$
calls to MPFUN subroutines, operating at multiple precision 1700.
This takes less than half an hour
on a DEC Alpha 3000-600 machine, corresponding to a call rate that is
faster than 1~kHz. The memory requirement is less than
1 MB: an array of $440\times240$ 4-byte cells holds the truncations in
multiple precision and is updated iteratively by\Eq{it}.

Of course, the above method {\em fails\/} as soon as one sets any of the
arguments to unity. However, all is not lost. By iteratively applying the
Euler-Maclaurin formula, one arrives at the conclusion
that the maximum power of $\ln n$ in the truncation error
$\sum_{i,j}A_{i,j}(\ln n)^i/n^j$ is the {\em largest\/} number, $d$, of
{\em successive\/} units in the string of arguments, since
only for $j=1$ does the integration of $(\ln x)^i/x^j$
increase the power of $\ln x$.
Consider, for example, $\zeta(a,1,b,1)$, with $a\neq1$ and $b\neq1$.
By the time the logarithm generated by the $n_4$ summation is felt by
the $n_2$ summation, it has acquired an inverse power of
$n_2$, thanks to the benignity of the Euler-Maclaurin formula
for the $n_3$ summation. Thus this sum has {\em demon\/}-number $d=1$,
even though it contains two unit arguments.
On the other hand, $\zeta(a,b,1,1)$ and $\zeta(a,1,1,b)$ have $d=2$,
as does $\zeta(a,1,1,b,1,a,1,1)$, for example.
At level $l$ the most demonic convergent sum has unity for all its arguments,
except for the first, which must therefore be $a_1=-1$. Fortunately,
it is possible to give exact expressions for this beast, with $d=l-1$,
and another, with $d=l-2$.
With a string of $n$ unit arguments denoted by $\us{n}$,
the all-level results
\begin{equation}
\zeta(-1,\us{l-1})=\frac{(-\ln2)^l}{l\,!}\,,\quad
-\zeta(-1,-1,\us{l-2})=\li{l}
=\sum_{n=1}^\infty\frac{1}{2^n n^l}\,,\label{ll}
\end{equation}
were inferred numerically, and then obtained analytically, by
Jon Borwein and David Bradley, as the coefficients of $t^l$
in $G(t,-1)$ and $t\int_0^1{\rm d}z\,G(t,z)/(1+z)$, generated by
the trivially summable hypergeometric series
$G(t,z)=t z\,{}_2F_1(1+t,1;2;z)=(1-z)^{-t}-1$. {From}
$t\int_0^1{\rm d}z\,G(t,z)/z$ one generates the corollary,
$\zeta(2,\us{l-2})=\zeta(l)$, of a theorem~\cite{AG}
\begin{equation}
\sum_{a_i>\delta_{i,1},\ l=\Sigma_i a_i}\zeta(a_1,a_2,\ldots,a_k)=
\zeta(l)\,,\label{AG}
\end{equation}
from Andrew Granville,
which was used to check evaluations of non-alternating sums.

To mitigate the computational difficulties caused by unit arguments,
truncation errors may be obtained analytically
for {\em combinations\/} of Euler sums of the form
\begin{equation}
S(a;b_1,\ldots,b_{k-1})
=\sum_{n=1}^\infty
\frac{({\rm sign}\,a)^{n}}{n^{|\,a|}}
\prod_{i=1}^{k-1}
\sum_{m_i=1}^{n-1}\frac{({\rm sign}\,b_i)^{m_i}}{m_i^{|\,b_i|}}\,.
\label{Salt}
\end{equation}
It seems appropriate to call\Eq{Salt} a {\em boxed\/} sum, since
the symmetrical inner summations span a lattice that is confined to a
$(k-1)$-dimensional hypercube by the outer summation variable.
It is built out of symmetrical combinations of Euler
sums with depths no greater than $k$, and is, so to speak, a
`cheap boxed set', available at a reduced~\cite{BaBG} computational price,
since it requires only the
{\em multiplication\/} of $k-1$ polygamma Euler-Maclaurin series,
followed by a {\em single\/} further application of the Euler-Maclaurin
formula, to determine the truncation error to any order for which ones
favorite computer-algebra engine has the power to multiply, differentiate, and
integrate double series in $\ln n$ and $1/n$. Only depth-1 data is needed,
since
the inner summations can be rewritten as
$\sum_{m_i}-\sum_{m_i\geq n}$, with the first term giving a constant
and the second an asymptotic series in $1/n$,
except for $b_i=1$, which gives Euler's constant, a log,
and an asymptotic series.

This Euler-Maclaurin method is the obvious generalization
of that used in~\cite{BaBG}, in the restricted cases
with $b_i=1$, or $b_i=-1$.
In the general case, much
computer-algebra time may be consumed by multiplying
asymptotic series for the distinct values of $b_i$ and then
integrating a long expression involving many powers of $\ln n$ and $1/n$,
though the algorithm is straightforward to implement.
To obtain a few hundred significant figures
in a short time one may stay within REDUCE, without feeding
thousands of lines of FORTRAN statements to TRANSMP;
at higher precision one benefits from this
transfer of function, at the expense of non-trivial file management,
when handling many boxed sums.

{From} an analytical point of view,
it is highly significant that every 3-fold Euler sum
can be transformed into a boxed sum:
$\zeta(a_1,a_2,a_3)\simeq-S(a_2;a_1,a_3)$,
where $\simeq$ will henceforward stand for
`equality modulo terms of lesser depth, their products,
and consequent questions of convergence'.
The argument is simple~\cite{BG}:
to the summation with $N\geq n_1>n_2>n_3\geq1$, for the truncation
of $\zeta(a_1,a_2,a_3)$, add that with $N\geq n_2>n_{1,3}\geq1$,
for the truncation of $S(a_2;a_1,a_3)$.
Adding the case $n_1=n_2$, which has lesser depth,
one has covered the values $N\geq n_1\geq1$, and hence has
a product of truncated sums of lesser depth.
Thus $\zeta(a_1,a_2,a_3)+S(a_2;a_1,a_3)\simeq0$.
The questions of convergence obviously concern
the case with $a_2=1$, in the limit $N\to\infty$.
Such issues are handled with great dexterity in~\cite{BG},
in the case of non-alternating triple sums.
Dealing, as now, with truncated sums, no problem of convergence
arises. The snag is that one must multiply truncations to
obtain the product term,
so the process becomes both messy and {\em ad hoc},
from the perspective of a programmer seeking a systematic
algorithm for sums of any depth. {From} a numerical point of
view, little is gained from boxability at depth $k=3$.

{\em Truly\/} nested Euler sums\Eq{zdef} have depth $k\geq4$, where
only combinations of them can be boxed. In the next subsection
it will be shown that at $k=4$ there first occurs a significant
phenomenon, of central relevance to the claim of Section~2, and to
knot-theoretical studies~\cite{DK1,DK2,AI95,BGK}.
But first, there is an outstanding computational
dilemma to confront. Should one attempt to automate the process of
{\em chaining\/} applications of the Euler-Maclaurin formula, feeding
in the required values for {\em multiple\/} Euler sums from runs of lesser
depth, and carefully separating the odd and even summations,
at each link of a chain whose length one would like to vary?
Or should one, rather, use the empirical truncation data to accelerate the
convergence? In the absence of logs, the one-line procedure\Eq{it}
settles the issue, to my mind. Why use masses of computer algebra to
feed into TRANSMP information that is already sitting in the numerical data,
in such immediately usable form?
When there are logs, from unit arguments, the matter is moot.
Having taken the empirical approach when the going was easy, I opted
to stick with it, when the going got tough, at depths $k\geq4$, with
$d>0$ demons.

It is clear that high-precision knowledge of $K(d,M)=1+(d+1)M$ truncated
values, for a sum with demon-number $d$,
should suffice to accelerate the convergence by a factor of roughly
$M!/N^M$, as was achieved with\Eq{it}, for $d=0$.
How best to achieve this, for $d>0$, is another matter. Unable to devise
an easily programmable iterative method like\Eq{it}, I returned
to the brute-force method of using $K(d,M)$ truncations to solve
directly for $K(d,M)$ unknowns, as was done a decade ago~\cite{5LB}
in the investigations which first suggested the appearance of
irreducibles with depth $k>1$
at the six-loop level of renormalization of quantum field theory,
a prediction amply confirmed by recent
six- and seven-loop analysis~\cite{AI95}, and illuminated by
knot theory~\cite{DK1,DK2}.

After translating the Gauss-Jordan~\cite{NR} method into MPFUN calls,
one may systematically obtain the $K(d,M)$ coefficients,
in the approximation
\begin{equation}
S(n)\approx S_\infty+\sum_{i=0}^d(\ln n)^i\sum_{j=C}^{C+M-1}
\frac{A_{i,j}}{n^j}\,,\label{pos}
\end{equation}
from $K(d,M)$ truncations, and in particular find an accurate value
for $S_\infty$. I chose the truncations $\{S(N+pL)\mid 0\leq p\leq(d+1)M\}$,
taken distances $L$ apart, starting at some value $N$. The choices
of $M$, $N$, $L$, and working accuracy, to achieve a desired output
accuracy for a sum with demon-number $d$, are an art learnt by experience,
not yet a science that is fit to be explained here. Suffice it to say that
it is wise to work at the {\em cube\/} of the desired accuracy,
and that solving
$10^3$ sets of $10^3$ equations,
at a working accuracy of one part in $10^{10^3}$,
in order to perform $10^3$ Euler-sum searches with
the MPPSLQ~\cite{MPPSLQ} integer-relation finder,
is demanding of core memory, CPUtime, vigilance, and patience.
The rewards, in increased {\em analytical\/} understanding, are considerable.

\subsection{Exact numerical results}

By an exact numerical result, I mean an equation whose exactness
is beyond intelligent doubt, yet is validated, to date,
only by {\em very\/} high precision numerical evaluation. It is, strictly
speaking, {\em possible\/} that an equation presented here is a divine hoax,
just as the rationality of $\zeta^2(3)/\zeta(6)$
is still a possibility. The improbability beggars all description.

The exact numerical result that sparked the genesis of Tables~1 and~3,
and revealed their utter simplicity,
was found at level $l=12$ and depth $k=4$. It reads
\begin{eqnarray}
&&2^5\tm3^3\,\zeta(4,4,2,2)=2^5\tm3^2\zf3
+2^6\tm3^3\tm5\tm13\ze9\ze3
+2^6\tm3^3\tm7\tm13\ze7\ze5\nl
+2^7\tm3^5\ze7\ze3\ze2
+2^6\tm3^5\zs5\ze2
-2^6\tm3^3\tm5\tm7\ze5\ze4\ze3\nl
-2^8\tm3^2\ze6\zs3
-\frac{13177\tm15991}{691}\ze{12}\nl
+2^4\tm3^3\tm5\tm7\zd62\ze4
-2^7\tm3^3\zd82\ze2
-2^6\tm3^2\tm11^2\zd{10}{2}\nl
+2^{14}\zu93\label{zu93}
\end{eqnarray}
and the sting is in its tail.

Exceptionally, factorizations of rationals were written in\Eq{zu93},
with a central dot (not to be confused with a decimal point)
denoting multiplication. If one dislikes the number 691,
one may remove it, using $\zeta(12)
=691\pi^{12}/(3^6\tm5^3\tm7^2\tm11\tm13)$.
Factorizations were given above lest the reader found the
unfactorized rationals implausible, on first encountering
an exact numerical result obtained by MPPSLQ.
The practice will not be continued.

It is apparent that one needs at least 100 significant figures
to discover\Eq{zu93}, because of the product of two 5-digit primes in
the $\pi^{12}$ term.
MPPSLQ (almost always) finds the integer relation,
$\sum_{i}n_i s_i\approx0$, with smallest euclidean norm,
$(\sum_i n_i^2)^{1/2}$, consistent with the requested accuracy of fit.
If one knew $\zeta(4,4,2,2)$ to only 100 significant
figures, the routine would be
at perfect liberty to return 13 `random' 8-digit integers that
happened to fit at 100 significant figures. It would not care that the true
form has attractive factorizations of all integers save one.
At 800 significant figures,
which is the accuracy to which\Eq{zu93} has been validated,
the probability of it being spurious is of order $10^{-700}$.

The import of\Eq{zu93} is dramatic: non-alternating sums, with
exclusively positive arguments in\Eq{zdef},
do {\em not\/} inhabit a cosy little world of their own,
uncontaminated by contact with their alternating cousins, as
the presence of $U_{9,3}$ clearly demonstrates.

As explained in the introduction, this was wonderful news for the connection
between knot theory and quantum field theory~\cite{DK1,DK2,AI95,BGK}.
It was also what
sparked the present systematic inquiry
into the {\em whole\/} universe of Euler sums of the form\Eq{zdef},
setting firmly aside the notion that there is something special about
non-alternating sums.
The liberating effect is apparent in the discovery of the simple
formul{\ae}\Eqq{Plk}{Gk}, which give order
to the larger universe so embraced.

Inspection of Table~3 reveals some practical limits to exploration.
Even with 800 significant figures
one might expect to encounter problems finding
relations in $S_{15,3}$, or $S_{10,4}$, each of which has 36 basis terms.
If just one of the 37 integers in a desired integer relation
exceeds $10^{23}$, then MPPSLQ may (quite properly)
fit 800-significant-figure data
with 37 `random' 22-digit integers.
Experience shows that integers of order $10^{22}$ are produced by
successful searches in $S_{13,3}$, with
28 basis terms. So investigation of $S_{15,3}$ was judged
to be imprudent, with `only' 800 significant figures at hand.
Because of the importance
of level 10 in~\cite{AI95}, investigation of $S_{10,4}$
is reported in Section~6.

The successful fit at all levels up to and including $l=7$,
using the 12 numbers\Eq{to7}, has been reported, as have the all-level
results\Eq{ll}. There remains the apparently trifling
matter of an exact
numerical relation in $S_{6,2}$:
\begin{equation}
\zu42=\df{97}{96}\ze6-\df34\zs3\,,\label{inn}
\end{equation}
with coefficients that Euler could, no doubt, have found by
mental arithmetic.
Surprisingly, nothing in the most recent work on double sums~\cite{BBG}
suggested the existence of such a relation.
It turned out to be relatively easy
to devise an analytical proof, when thus apprised of its need,
so the subject is postponed to the next section,
where the armory of analytical tools is augmented. Yet
the relation belongs here, since it was
David Bailey's engine that disclosed it.
The analytical techniques, developed to derive\Eq{inn}, feed back useful
information for further numerical analysis.

First, they clear up the whole of the
double-sum sector, for good and all, confirming the
knot-theoretic expectation
of a `rule of 2', with $E(2p,2)=\ip{p/2}$, and thus
allowing numerical exploration to progress to $k\geq3$.
Secondly, they provide rigorous (though non-optimal) upper bounds:
\begin{equation}
E(7,3)\leq3\,,\quad
E(8,4)\leq4\,,\quad
E(9,3)\leq6\,,\quad
E(11,3)\leq11\,,\quad
E(13,3)\leq17\,.
\label{bound}
\end{equation}
Thirdly, for each bound it is possible to find
an overcomplete basis, whose size is determined by the bound.
Thus one needs to evaluate only a small fraction of the
sums in these sectors, and then use MPPSLQ to reduce an overcomplete
basis to a minimal basis of size $S(l,k)$.
If the formula for $S(l,k)$ were false, MPPSLQ might
sometimes fail to reduce the basis down to the claimed size, or might
reduce it down to a size smaller than that claimed by\Eq{Slk}. Of course,
it never did the latter, else the claim would not have been made.
The fact that it never did the former, with 800-significant-figure
data, in search spaces
of sizes up to 36, is a testament to its author~\cite{MPPSLQ} as well
as to the formula.
Finally, and most fortunately, it is possible to construct overcomplete
bases, in the spaces bounded by\Eq{bound},
that are demon-free. Hence 800 significant figures are available,
in less than half an hour per sum, at the touch of button\Eq{it}.

Thanks to these circumstances, all Euler sums in
$\{S_{l,3}\mid l\leq14\}$ and $\{S_{l,4}\mid l\leq9\}$ have been
shown to be reducible to bases of the sizes given in Table~3,
by the operation of MPPSLQ on overcomplete bases.
Relatively simple examples of such reductions,
in $S_{9,3}$, $S_{11,3}$ and $S_{8,4}$, are provided by:
\begin{eqnarray}
\zo333&=&
 6\zo513
+6\zo315
-\df{315}{32}\lt\ze3\ze5
+6\zu51\ze3\nq
&+&\df{40005}{128}\ze2\ze7
-\df{39}{64}\zc3
+\df{1993}{256}\ze3\ze6
+\df{8295}{128}\ze4\ze5
-\df{226369}{384}\ze9,\nonumber\\[3pt]
\zo353&=&
 \df{1059}{80}\zn533
+15\zo713
+15\zo317
+\df{701}{69}\zu53\ze3\nq
&+&15\zu71\ze3
-\df{6615}{256}\lt\ze3\ze7
-\df{11852967}{2560}\ze{11}
+\df{301599}{128}\ze2\ze9\nq
&-&\df{124943}{5888}\zs3\ze5
+\df{1753577}{35328}\ze3\ze8
+\df{2960103}{5120}\ze4\ze7
+\df{3405}{32}\ze5\ze6,\nonumber\\[3pt]
\!\zeta(3,-1,3,-1)&=&
\df{61}{27}\,\zeta(-3,-3,-1,-1)
-\df{14}{3}\,\zeta(-5,-1,-1,-1)
-\df{185}{27}\zu51\ze2\nq
&-&\df{163499}{22356}\zu53
+\df{2051}{54}\zu71
+\df{28}{9}\lp2\zu51
+\df{35}{96}\lp2\zs3\nq
&-&\df{581}{64}\lp2\ze6
-\df{8735}{576}\lt\ze2\ze5
-\df{903}{64}\lt\ze3\ze4\nq
&-&\df{1441}{288}\ze2\zs3
+\df{10365875}{476928}\ze3\ze5
+\df{36916435}{1907712}\ze8.\label{cnot}
\end{eqnarray}

In $S_{13,3}$, the relations are
more complex, with the prime factor 102149068537421 appearing in one case.
Nonetheless, the probability of a spurious fit is less than $10^{-200}$,
in all cases, and is often much less than this.
The existence of further relations, forbidden by\Eq{Gk},
cannot be disproved by numerical methods.
The euclidean norms of such unwanted relations would, however,
exceed those of the discovered
relations by factors ranging between $10^{10}$ and $10^{20}$, which
makes it {\em extremely\/} implausible that the formula is in error in
any of the spaces with $S(l,k)\leq S(8,8)=F_9=34$.

\section{Posterior analytics}

\subsection{Analytical tools}

The analysis of~\cite{BBG,BG} makes use of two very simple types
of relation between Euler sums. In the general case of $k$-fold sums,
with all possible alternations of sign,
it is somewhat difficult to notate these relations,
in all generality. To avoid cumbersome formul{\ae}, terms that involve
sums of lower depth, and their products, will be omitted, as in the
case of $\zeta(a_1,a_2,a_3)\simeq-S(a_2;a_1,a_3)$.

The first type of relation involves permutations of arguments:
\begin{eqnarray}
0&\simeq&\zeta(a_1,a_2,a_3,\ldots,a_k)
+\zeta(a_2,a_1,a_3,\ldots,a_k)
+\zeta(a_2,a_3,a_1,\ldots,a_k)
+\ldots\nonumber\\
&+&\zeta(a_2,a_3,a_4,\ldots,a_1)\,.
\label{perm}
\end{eqnarray}
The proof is trivial: by including all insertions of $a_1$ in the
string $a_2,a_3,\ldots,a_k$, one obtains a combination of sums
that differs from the product $\zeta(a_1)\,\zeta(a_2,a_3,\ldots,a_k)$
only by terms in which summation variables are equal, corresponding
to sums of depth $k-1$.
For 4-fold sums, such
relations reduce a set of 24 possible permutations to a set of 9,
when the arguments are distinct.

The second type of relation follows from use of the
partial-fraction identity~\cite{BBG,BG}
\begin{equation}
\frac{1}{A^a B^b}=\sum_{s>0}\frac{1}{(A+B)^{a+b-s}}
\left\{{a+b-s-1\choose a-s}\frac{1}{A^s}+{a+b-s-1\choose b-s}\frac{1}{B^s}
\right\}\,
\label{ab}
\end{equation}
for positive integer $a$ and $b$. To see how this is used, consider
the product $\zeta(a)\zeta(b,c,d)$, with positive arguments. It may be
written as $\sum_{n,p,q,r} 1/n^a(p+q+r)^b(q+r)^c r^d$,
where each summation variable runs over all the positive
integers. Now apply\Eq{ab}, setting $A=n$ and $B=p+q+r$.
The second type of resulting
partial fraction is of the form $\sum_{n,p,q,r}
1/(n+p+q+r)^{a+b-s}(p+q+r)^s(q+r)^c r^d$,
which is an Euler sum. To the first type,
apply\Eq{ab} with $A=n$ and $B=q+r$.
Its second terms are also Euler sums. To its first, apply\Eq{ab} with
$A=n$ and $B=r$. Each term so produced is an Euler sum.
Thus one has obtained a relation for non-alternating sums. By including
signs, 16 such relations can be generated. In general,
one gets $2^k$ relations by $k-1$
applications of\Eq{ab} for every set of $k$ exponents that
one chooses for the initial product of sums.

It can seen that there is no scarcity of trivially derivable relations
between Euler sums. The notable achievement of~\cite{BBG} was to organize the
relations between double sums in such a way as to prove the reducibility
of all double sums of odd level. In~\cite{BG} non-alternating triple
sums of even level were proven to be reducible. It was conjectured that
non-alternating sums of level $l$ and depth $k$ are reducible whenever
$l+k$ is odd. The stronger claim made by\Eq{Elk} is that this applies
to alternating sums as well. In the course of the present work,
reducibility has been demonstrated, by a combination
of analytical and numerical methods, for all odd values
of $l+k$ such that $S(l,k)\leq S(14,3)=29$.

As remarked previously, the identities of~\cite{BBG} are insufficient to
derive the simple relation\Eq{inn}. No tally was given in~\cite{BBG}
of the numbers of alternating double sums left unreduced at even levels,
though the tally $\ip{p/3}$ was made for non-alternating double sums at
level $2p+2$. Using REDUCE, one easily discovers
that the relations given in~\cite{BBG} allow reduction of double
sums to the set $\{\zu{n+2m}{n}\mid\min(n,m)>0\}$ and that no further
reduction is possible without additional input. Using MPPSLQ, one easily
discovers that a truly irreducible set is furnished by
$\{\zu{2a+3}{2b+1}\mid a\geq b\geq0\}$.
Thus the relations derived in~\cite{BBG} are
insufficient in a way that is very easy to state: they fail to relate
the even cases
$\{\zu{2a+4}{2b+2}\mid a\geq b\geq0\}$ to the odd cases
$\{\zu{2a+3}{2b+1}\mid a\geq b\geq0\}$.

To remedy this failure, it was sufficient to derive the
further\footnote{Jon Borwein later told me that\Eqqq{zab}{za}{zb}
were known to,
though not used by, the authors of~\cite{BBG}.} relation
\begin{eqnarray}
\zeta(a,b)+\zeta(-a,-b)&=&\sum_{s>0}(a+b-s-1)!\left\{
 \frac{\zeta_A(a+b-s,s)}{(a-s)!\,(b-1)!}
+\frac{\zeta_B(a+b-s,s)}{(b-s)!\,(a-1)!}
\right\}\label{zab}\\
\zeta_A(a,b)&=&\zeta(a,b)+\zeta(-a,b)
-2^{1-a}\left\{\zeta(a,b)+\zeta(a+b)\right\}\label{za}\\
\zeta_B(a,b)&=&2^{1-a}\zeta(a,b)\label{zb}
\end{eqnarray}
with $a>1$ and $b>1$. To prove\Eq{zab}, one writes
$\zeta(a,b)+\zeta(-a,-b)=\sum_{m,n} 2/(2m+n)^a n^b$.
Then\Eq{ab}, with $A=2m+n$ and $B=n$, yields\Eqq{za}{zb},
after some rearrangements.

\subsection{Double sums in knot theory and field theory}

Adjoining\Eq{zab} to relations in~\cite{BBG}, it was possible
to use REDUCE to derive expressions
for the 3698 double sums up to level 44, in terms of the
121 irreducible double sums
\begin{equation}
\{\zu{2a+3}{2b+1}\mid a\geq b\geq0,\ a+b \leq 20\}\,.\label{done}
\end{equation}
The family of positive knots~\cite{BGK}
that gave rise to this investigation has
braid\footnote{For an introduction to knot theory,
try~\cite{CCA}, followed by~\cite{DR,VFRJ}.}
words
\begin{equation}
\{\fk{}{2a+1}{}{2b+1}\mid a\geq b\geq1\}\,,
\label{braid}
\end{equation}
whose enumeration satisfyingly matches that of the irreducibles\Eq{done}.

Note that one omits the knots
$\{\fk{}{2a+1}{}{}\mid a\geq0\}$ from the
tally of 3-braids in\Eq{braid},
since Reidermeister moves
transform them to $\{\so{2a+3}\mid a\geq0\}$,
which are the 2-braid torus knots,
corresponding~\cite{DK2} to the depth-1 irreducibles
$\{\zeta(2a+3)\mid a\geq0\}$, whose occurrence has
been studied to all~\cite{BU,eva,UD,lad} orders in
quantum field theory.
Correspondingly, the Euler sums
$\{\zu{2a+3}{1}\mid a\geq0\}$ do not occur in
counterterms\footnote{For details of renormalization procedures,
see~\cite{VAS}, pending publication of~\cite{DK}.}
though they may appear in the finite parts
of integrals obtained from Feynman diagrams, and hence in
the relationships between physical quantities, such as the charge
and magnetic moment of the electron.

In fact, two of the most impressive perturbative calculations~\cite{AFMT,LR}
of physical quantities
in quantum field theory produce the polylogarithm $\li4$,
with the precise combination of $(\ln 2)^4$ and $\pi^2(\ln 2)^2$ terms
given in\Eq{U31} for $U_{3,1}=\frac12\ze4-2\al4$. The
$\rho$\/-parameter~\cite{AFMT} of electroweak theory entails,
at three loops, the level-4 term of~\cite{WP2}
\begin{equation}
B_4=-\left\{8\zu31+\df52\ze4\right\}+{\rm O}(\ep)\label{B40}
\end{equation}
in $4-2\ep$ spacetime dimensions. The level-4 terms
in the three-loop contributions to the anomalous magnetic moment
of the electron, $\frac12(g-2)_{\rm e}$,
are obtained from~\cite{LR} as
\begin{equation}
-\left\{\df{50}{3}\zu31+\df{13}{8}\ze4\right\}
(e/2\pi)^6\,,\label{g3}
\end{equation}
where $-e$ is the electron's charge, in units of $(\ep_0\hbar c)^{1/2}$.
Thus the lowest-level double-sum irreducible, $U_{3,1}$,
is prominent in quantum field theory, though absent
from counterterms.

Having seen the importance of double Euler sums in quantum field
theory, and their relation to knot theory, one should move on to 3-fold
alternating sums, since these too occur in field theory, as will be shown
in Section~7. Unfortunately, the tools are not yet available to
derive analytically {\em all} the relations implied by\Eq{Gk}, with $k>2$.
One must, therefore, make do with analytical derivations of
{\em most} of them.

\subsection{Proven bounds}

At depths 3 and 4, the rigorous bounds\Eq{bound} were proven by
implementing the
permutation\Eq{perm} and partial-fraction\Eq{ab} procedures in
REDUCE, using its solve command. The results are conveniently
returned in terms of ARBCOMPLEX~\cite{RED}
variables equal in number to the
undetermined sums. One may then use the output to form a proven,
overcomplete, demon-free basis. This was achieved for all triple sums up to
level 14, and all quadruple sums up to level 9, which is no small
undertaking, as may be judged from the facts that two days of CPUtime were
insufficient to solve for the 676 triple sums at level 15, and 128
megabytes of core memory were insufficient to solve for the 1120 quadruple
sums at level 10. I recommend these sectors as test grounds for improved
algorithms.

In the case of triple sums up to level 14, overcomplete
bases were constructed by adjoining
the non-alternating~\cite{BG} irreducibles
$\{\zeta(5,3,3),\zeta(7,3,3),\zeta(5,5,3)\}$
to the set
\begin{equation}
\{\zeta(2n+3,-2m-1,-2p-1)\mid n+m+p\leq 4,\ \min(n,m,p)\geq0\}\,,\label{toc}
\end{equation}
which is also undetermined by the permutation and partial-fraction identities.
Attempts to reduce this set analytically, by
adding identities obtained in the manner of\Eq{zab}, were not
successful. Therefore the most pressing challenge is to {\em prove\/}
the MPPSLQ result that $E(7,3)\leq2$, since only the bound $E(7,3)\leq3$
has been obtained rigorously, from\Eq{toc}.
In the case of quadruple sums up to level 9, the analytical method fails to
find two MPPSLQ relations in $S_{8,4}$, one of which is given in\Eq{cnot}.
Hence I recommend the $S_{7,3}$ and $S_{8,4}$ sectors as places
to start the hunt for more powerful analytical techniques.

\section{Concrete bases}

To aid the elucidation of\Eqq{Elk}{Plk} in Section~8.3,
it is instructive to summarize the results of MPPSLQ, from Section~4,
and REDUCE, from Section~5, by giving concrete irreducibles
whose values and products span, {\em inter alia\/}, all those spaces of
Table~3 with $S(l,k)<36$. Also included is a choice of the 5 irreducibles
in $S_{10,4}$, which proved attainable with MPPSLQ, despite the large size,
$S(10,4)=36$. To save space, only argument strings are given, with
a bar denoting a negative argument, and hence an alternation of sign
at the corresponding layer of the nest.

For $k=1$, one obviously needs $\ln2$, $\pi^2$, and
the odd-zetas, with argument strings $\{(2n+1)\mid n>0\}$.
For $k=2$, the argument strings $\{(\ovl{2n+1},\ovl{2m+1})\mid n>m\geq0\}$
suffice, to level $l=44$, and presumably for ever.
For the remaining spaces with $S(l,k)<36$, see Table~4, which also
includes $S_{10,4}$.

{\bf Table~4}:
Concrete sets of argument strings,
yielding demon-free minimal bases.
\[\begin{array}{llllllll}
S_{5,3} & \bt311   \\
S_{7,3} & \bt511   & \bt313 \\
S_{9,3} & \bt711   & \bt513 &\bt315 \\
S_{11,3}& \bt911   & \bt713 &\bt515 &\bt533 & \bt317 \\
S_{13,3}& \bt{11}11& \bt913 &\bt715 &\bt733 & \bt517 &\bt535 &\bt319 \\[3pt]
S_{6,4} & \bq3111  \\
S_{8,4} & \bq5111  & \bq3311\\
S_{10,4}& \bq7111  & \bq3331&\bq5311&\bq5131&\bq3151 \\[3pt]
S_{7,5} & (3,\ovl1,\ovl1,\ovl1,\ovl1)\\[3pt]
S_{8,6} & (\ovl3,\ovl1,\ovl1,\ovl1,\ovl1,\ovl1)
\end{array}\]

It is emphasized that the choices of Table~4 are {\em not\/} the most
efficient for integer-relation searches; other choices, as in\Eq{to7},
result in smaller euclidean norms from successful searches. However, one
learns more from the patterns of Table~4 than from the economic choices
of\Eq{to7}.

First, note that only odd arguments are required: the pattern
begun by odd-zetas at depth 1 persists.
Secondly, note that all arguments may taken as negative,
save the first, when the depth, $k$, is odd.
Since the first argument cannot be unity,
demons are thereby eliminated from the irreducibles.
Each is thus computable, to 800 significant figures, in half an hour,
using\Eq{it}.
Finally, note that not all
permutations of odd integers occur.

Whether these features persist in larger search spaces,
where MPPSLQ may well need more than 800 significant figures,
can only be conjectured. The argument of Section~8.3, however,
lends weight to the belief that the key to understanding irreducibility,
at level $l$ and depth $k$, lies in an analysis of a {\em restricted\/} set of
permutations of partitions of $l$ into precisely $k$
positive {\em odd\/} integers.

\section{Euler sums in quantum electrodynamics}

To exemplify the utility and accuracy of the database of exact result for
all sums with $l\leq7$, it was used to evaluate the Laurent expansion of
the three-loop on-shell charge-renormalization~\cite{WP2} constant of
dimensionally regularized~\cite{tHV} quantum electrodynamics, up to terms of
level 7, which may be taken as indicative of the transcendentality content
of four-loop contributions to the anomalous magnetic moment of the
electron.

Like the three-loop corrections~\cite{AFMT}
to the $\rho$\/-parameter of electroweak
theory, on-shell charge renormalization
involves~\cite{WP2}
\begin{eqnarray}
B_4&=&
\frac{(\mu-1)(\mu-\frac32)}{\pi^{3\mu}\Gamma^3(3-\mu)}\int\int\int
\frac{{\rm d}^{2\mu}p\,{\rm d}^{2\mu}q\,{\rm d}^{2\mu}r}
{(p^2+1)(q^2+1)(r^2+1)(p-q)^2(p-r)^2}\nl
\phantom{\frac{(\mu-1)(\mu-\frac32)}{\pi^{3\mu}\Gamma^3(3-\mu)}\int\int\int}
\times\left\{\frac{1}{(p-q-r)^2+1}-\frac{1}{(q-r)^2}\right\}\,,
\label{B4}
\end{eqnarray}
which is a difference of three-loop massive bubble
diagrams, regularized in $2\mu\equiv4-2\ep$ euclidean
spacetime dimensions.
In~\cite{WP2} it was reduced to a ${}_3F_2$ hypergeometric
series~\cite{GHH}:
\begin{eqnarray}
B_4&=&
\frac{7}{24\ep^4}\left\{1-\frac{\Gamma(1-\ep)
\,\Gamma^2(1+2\ep)\,\Gamma(1+3\ep)}{\Gamma^2(1+\ep)\,\Gamma(1+4\ep)}\right\}
-\frac{\pi^2}{3\ep^2}\,\frac{\Gamma(1+2\ep)\,\Gamma(1+3\ep)}
{2^{6\ep}\,\Gamma^5(1+\ep)}\nl
+\frac{8}{3\ep^2(1+2\ep)}\,{}_3F_2\!\left(1,\df12-\ep,\df12-\ep;
\df32+\ep,\df32;1\right)\,.\label{F32}
\end{eqnarray}
Expanding the $\Gamma$\/-functions in the summand of the series,
one obtains the $\ep$\/-expansion of $B_4$ from boxed sums of the form
\begin{equation}
S_{\rm odd}(a;b_1,\ldots,b_{k-1})
=\sum_{n=1}^\infty\frac{1}{(2n-1)^a}
\prod_{i=1}^{k-1}\sum_{m_i=1}^{n-1}\frac{1}{(2m_i-1)^{b_i}}\,,
\label{Sodd}
\end{equation}
which, like\Eq{Salt}, is symmetric in $\{b_i\}$, though now
it involves reciprocal powers of {\em odd\/} integers.
To relate these to\Eq{zdef}, one combines $2^k$
$k$\/-fold Euler sums with arguments of differing sign, to restrict the
nested summations to odd integers.
Symmetrizing over $(k-1)!$ permutations of
all but the first argument, one exhausts those summations
in\Eq{Sodd} with distinct values of $\{m_i\}$.
Adding combinations of Euler sums with lesser depth, one iteratively
includes the degenerate summations. Hence
the expansion of $B_4$ to ${\rm O}(\ep^3)$ can be achieved in terms of
the basis\Eq{to7}, by routine (and repeated) reference to the database.
The result is
\begin{eqnarray}
B_4&=&
\df12\left\{
-13\ze4+32\al4
\right\}
+\df12\ep\left\{
-239\ze5+192\al5+204\lt\ze4
\right\}\nl
+\df32\ep^2\left\{
13\ze6-74\zs3+160\zu51+384\al6-204\lp2\ze4
\right\}\nl
+\df{1}{72}\ep^3\left\{
-329385\ze7+45853\ze3\ze4+10875\ze2\ze5+7680\zt51
\right.\nl\left.
-9280\zt33-11520\al4\ze3+176031\lt\ze6-2930\lt\zs3
\right.\nl\left.
-48000\lt\zu51+248832\al7+44064\lp3\ze4
\right\}+{\rm O}(\ep^4)\,.
\label{B4e}
\end{eqnarray}
Note that the polylogarithms\Eq{alpha} simplify the expansion;
had one merely used the conventional~\cite{Lew}
polylogarithms $\li{n}$, there would have
been 10 additional terms, to this order in the expansion.
The terms involving $\lp{l-4}\ze4$ follow $\alpha(l)$,
being generated by
\begin{equation}
\sum_{l>4}(6\ep)^{l-4}\left\{16\al{l}
-\frac{17}{(l-4)!}(-\ln2)^{l-4}\ze4\right\}\,.\label{fol}
\end{equation}

A strong check of\Eq{B4e} was performed, using the representation
\begin{equation}
(1+2\ep)\left[\frac{\Gamma(1-\ep)}{\Gamma(1+\ep)\,\Gamma(1-2\ep)}\right]^2
\int_0^1\int_0^1\frac{{\rm d}x\,{\rm d}y}{1-x^2y^2}
\left[\frac{(1-x^2)(1-y^2)^2}{(8x y)^2}\right]^\ep\label{NAG}
\end{equation}
of the hypergeometric series
in\Eq{F32}, expanding the integrand to ${\rm O}(\ep^5)$, and
using the NAG routine D01FCF to evaluate 6 double integrals
to 8 significant figures.
By contrast, 800 significant figures were rapidly obtained by
expressing sums of the form\Eq{Sodd} in terms
of those in the database. The agreement of NAG with the more powerful
methods developed here gives one confidence in the
computer-algebraic book-keeping that produced\Eq{B4e}.

There is a further~\cite{mas} non-trivial diagram
entailed in three-loop charge renormalization:
the two-loop fermion-propagator~\cite{MZ1,MZ2}
diagram, with a
three-fermion intermediate state, contributing to
on-shell mass renormalization~\cite{MZ1,mat}.
However, this is eventually expressible~\cite{WP2} in
terms of $\Gamma$\/-functions
and the same hypergeometric series as is encountered in\Eq{F32}.
Thus no further analysis of irreducible Euler sums is entailed,
though there is a great deal of book-keeping to perform, to obtain
the three-loop terms in the dimensionally regularized
charge-renormalization constant~\cite{WP2}
\begin{equation}
Z_3=1+\sum_{n=1}^\infty C_n
\left(\frac{e^2\Gamma(1+\ep)}{(4\pi)^{2-\ep}m^{2\ep}}\right)^n\,,\label{Z3}
\end{equation}
where $e$ and $m$ are the on-shell charge and mass.

The one-loop and two-loop contributions are easily found exactly~\cite{MZ2}:
\begin{equation}
C_1=-\frac{4}{3\ep},\,\quad\
C_2=-\frac{4(1+7\ep-4\ep^3)}{\ep(2-\ep)(1-2\ep)(1+2\ep)}\,.\label{C12}
\end{equation}
The first three~\cite{WP2} terms in the Laurent expansion
\begin{equation}
C_3=-\frac{8}{9\ep^2}+\frac{62}{27\ep}+\left\{
128\lt\ze2
-{22\over3}\ze3
-{368\over3}\ze2
+{4867\over81}
\right\}
+\sum_{n=1}^\infty C_{3,n}\ep^n\label{cex}
\end{equation}
are free of contributions from $B_4$, whose expansion starts at level
$l=4$. Level-3 terms are sufficient for the analysis~\cite{BKT,BB,PAB} of a
restricted set of four-loop contributions to the anomalous magnetic moment
of the muon~\cite{TK}, and for the study of quark-mass effects at the
three-loop level of quantum chromodynamics~\cite{CKS}. Thanks to\Eq{B4e},
the expansion\Eq{cex} may now be continued to level 7, i.e.\ to the same
level as is reached at four loops in the electron's anomaly, whose
polylogarithms attain order $2L-1$ at $L$ loops. After much computer
algebra, I was able to obtain:
\begin{eqnarray}
C_{3,1}&=&
-384\lp2\ze2
+{1358\over3}\ze4
-{736\over3}\al4
+1024\lt\ze2
-{2957\over6}\ze3\nl
-{1960\over3}\ze2
+{104113\over486}
\cis{-179.
724\,615\,842\,918\,120\,241\,823\,332\,320\,650\,562\,692\,071\,547\,121}
1\,,\\
C_{3,2}&=&
768\lp3\ze2
-284\lt\ze4
+384\ze2\ze3
+{5101\over3}\ze5
-1472\al5\nl
-3072\lp2\ze2
+{55571\over18}\ze4
-{30488\over9}\al4
+4096\lt\ze2\nl
-{34537\over12}\ze3
-{7324\over3}\ze2
+{1937227\over2916}
\cis{-427.
138\,027\,736\,892\,466\,683\,630\,594\,488\,509\,635\,499\,554\,227\,264}
2\,,\\
C_{3,3}&=&
-1152\lp4\ze2
+852\lp2\ze4
-1280\lt\ze2\ze3
+{5018\over3}\zs3
+6356\ze6\nl
-8832\al6
-3680\zu51
+6144\lp3\ze2
-{34067\over3}\lt\ze4
+{6272\over3}\ze2\ze3\nl
+{591473\over36}\ze5
-{60976\over3}\al5
-12288\lp2\ze2
+{1472549\over108}\ze4\nl
-{450388\over27}\al4
+13696\lt\ze2
-{733013\over72}\ze3
-{25226\over3}\ze2
+{33051769\over17496}
\cis{-1371.
792\,496\,978\,355\,362\,371\,049\,120\,514\,541\,715\,942\,648\,466\,341}
3\,,\\
C_{3,4}&=&
{6912\over5}\lp5\ze2
-1704\lp3\ze4
+3840\lp2\ze2\ze3
+{33695\over54}\lt\zs3\nl
-{334273\over12}\lt\ze6
+{92000\over9}\lt\zu51
+{7360\over3}\al4\ze3
+{68689\over36}\ze2\ze5\nl
-{334907\over108}\ze3\ze4
+{2459549\over36}\ze7
-52992\al7
-{14720\over9}\zt51\nl
+{53360\over27}\zt33
-9216\lp4\ze2
+34067\lp2\ze4
-10240\lt\ze2\ze3\nl
+{129179\over6}\zs3
+33767\ze6
-121952\al6
-{152440\over3}\zu51\nl
+24576\lp3\ze2
-{1176869\over18}\lt\ze4
+{21184\over3}\ze2\ze3
+{15720695\over216}\ze5\nl
-{900776\over9}\al5
-41088\lp2\ze2
+{27328097\over648}\ze4
-{4616354\over81}\al4\nl
+43712\lt\ze2
-{14076461\over432}\ze3
-{82735\over3}\ze2
+{555842827\over104976}
\cis{-3273.
919\,335\,883\,520\,406\,469\,573\,320\,145\,714\,810\,021\,184\,454\,681}
4\,.
\end{eqnarray}

Terms\Eqq{C31}{C32} are in agreement with the polylogarithms of order up to
5 that were obtained in~\cite{WP2}, though they are much simpler in form,
thanks to the use of\Eq{alpha}. Terms\Eqq{C33}{C34} are new, and show how
complex perturbation expansions may become, when they
entail polylogarithms of orders 6 and 7, as undoubtedly happens in the
four-loop contributions to the electron's anomaly. However, I would be
rather surprised were the four-loop anomaly to contain {\em further\/}
transcendentals, not included above, since Euler sums are the natural
structures to emerge from $\ep$\/-expansions of generalized hypergeometric
series~\cite{GHH,WNB,LJS} whose parameters differ from
half-integers~\cite{WP2,BFT,GG2}, or integers~\cite{BGK,SJH,AVK}, by
multiples of $\ep$.  In fact, there is good reason to suppose that not all
of the terms above will occur in the four-loop anomaly, since Laporta and
Remiddi~\cite{LR} have shown that $\li5$ is absent at three loops.

\section{Consequences}

The enumerations\Eqqq{Elk}{El}{Plk} have consequences for field theory,
knot theory, and number theory, which will be discussed in that order.

\subsection{Field theory}

For calculational quantum field theorists, the enumeration\Eq{El}
amounts to `counting the enemy', since each new irreducible Euler sum
corresponds to the existence of a class of polylogarithmic integrals
that cannot
be related to previously evaluated integrals, by computer-algebraic methods.
They are few in number, at the levels where calculations are
currently performed. Their
tally, in Table~2, is the integer sequence M0317 of~\cite{SP}.

As reviewed in~\cite{q92,a92,a93}, there has been tremendous progress in
the use of computer algebra, most notably by recursive methods within the
framework of dimensional regularization, which automate the computation of
single-scale massless~\cite{CT,FVT} and on-shell massive~\cite{LR,WP2,PAB}
propagator diagrams by symbolic manipulation of vast~\cite{LRV} numbers of
polynomials in the analytically continued~\cite{tHV} dimensionality,
$2\mu=4-2\ep$, of spacetime. The residual difficulty then resides in
extracting the Laurent expansions, as $\ep\to0$, of a small set of
irreducible Feynman integrals. Recent success at three-loops~\cite{LR},
with the electron's anomalous magnetic moment, provides an impressive
example of the power of this technique. Following discussions at the
AI-HEP-92 workshop~\cite{a92}, Laporta and Remiddi found it possible to
achieve a computer-algebraic reduction of all 3-loop electron-anomaly
diagrams to merely 18 terms~\cite{LR}, using integration by
parts~\cite{CT,FVT} in $4-2\ep$ dimensions. Drawing on experience of
polylogarithmic integration, from previous 4-dimensional work, they were
able to extract the requisite Laurent expansions as $\ep\to0$. Along with
the inevitable depth-1 irreducibles, $\ln2$ and $\{\ze{n}\mid n\leq5\}$,
their result contains just one irreducible double Euler sum, namely
$U_{3,1}$, though it is presented in~\cite{LR} in terms of the more
conventional~\cite{Lew} polylogarithm $\li4$, along with the precise $(\ln
2)^4$ and $\pi^2(\ln2 )^2$ terms of\Eq{U31}.

In the case of field-theory counterterms, up to $L$ loops,
the irreducible Euler sums
\begin{eqnarray}
&&\{\ze{2a+1}\mid L-2\geq a>0\}\,,\label{2b}\\
&&\{\zu{2a+1}{2b+1}\mid a>b>0,\ L-3\geq a+b\}\,,\label{3b}\\
&&\{\zn{2a+1}{2b+1}{2c+1}\mid a\geq b\geq c>0,\ a>c,\ L-3\geq a+b+c\}
\label{4b}\,,
\end{eqnarray}
appear.
The first set, studied in~\cite{BU,eva,lad},
corresponds~\cite{DK1} to 2-braid torus knots, with the trefoil
knot $3_1\simeq\ze3$ first appearing in counterterms at $L=3$ loops.
The second~\cite{exp,zs6} set
corresponds~\cite{BGK} to a restricted set of positive 3-braids,
with $8_{19}\simeq\zu53$ first appearing at $L=6$ loops,
and the third~\cite{BG,AI95} to positive 4-braids,
with the uniquely positive 11-crossing
non-torus knot
$\eltft\simeq\zn533$
first appearing at $L=7$ loops. Hence the 5-loop renormalization
of $\phi^4$ theory was accomplished~\cite{phi}
in terms of only depth-1 irreducibles, whereas
at 6 and 7 loops one first encounters~\cite{AI95} 2-fold and 3-fold
Euler sums, respectively.

It is not yet known whether Euler sums exhaust the
transcendentals in counterterms at $L\geq7$ loops. In addition to
$10_{124}=\fk{}5{}3
\simeq\zu73$, there are two further
positive knots with 10 crossings, namely~\cite{VFRJ}
$10_{139}=\fk{}333$ and
$10_{152}=\fk2233$, which
7-loop analysis~\cite{AI95} suggests are {\em not}
associated with Euler sums of depth $k<4$. All those, and {\em only\/} those,
subdivergence-free 7-loop $\phi^4$ diagrams whose link diagrams skein
to $10_{139}$ and $10_{152}$ appear, on the basis of numerical evidence,
to give counterterms that cannot be reduced to Euler sums
with depth $k\leq3$. Hence it is an open (and fascinating) question
whether
two\footnote{Recall that $U_{3,1}$ and $U_{5,1}$ are absent from
counterterms; no positive knot has crossing number 4 or 6.}
of the 5 new irreducibles in $S_{10,4}$ are associated
with these knots.

Turning to the finite parts of Feynman diagrams, one learns from three-loop
analyses~\cite{AFMT,LR} that $U_{3,1}$ appears, via\Eqq{B40}{g3}.
Emphatically {\em no\/} claim is made that Euler sums exhaust the
transcendentality content of perturbative quantum field theory;
polylogarithms of non-trivial mass and momentum ratios are everywhere dense.
In single-scale process, however, where such ratios are unity, or zero, it
{\em may\/} occur that the results entail only Euler sums, as in
the case of electron's anomaly, $\frac12(g-2)_{\rm e}$
at three loops~\cite{LR}.
On the other hand, with a different configuration of unit and zero masses
the maximum value of Clausen's integral~\cite{Lew} is often
generated~\cite{mas,BFT}.

The key to deciding whether a result is reducible to Euler sums is an
analysis of the hypergeometric functions~\cite{BGK,WP2,BFT}
${}_{p+1}F_{p}(1,\{a_i\};\{b_i\};z)$ that are produced in $4-2\ep$
dimensions. If $z=\pm1$, and $\{a_i,b_i\mid i\leq p\}$ differ from
integers, or half integers, by multiples of $\ep$, then reducibility to
Euler sums is guaranteed, as in\Eqq{F32}{B4e}. Such a reduction to
hypergeometric series has been achieved for the electron anomaly,
$\frac12(g-2)_{\rm e}$, at two loops~\cite{WP2,FT}, for the
charge-renormalization constant, $Z_3$, at three loops~\cite{WP2}, for the
Gell-Mann--Low function of quantum electrodynamics to all orders in the
large-$N$ limit~\cite{lnf}, and for a corresponding limit of the quantum
chromodynamics of heavy-quark interactions~\cite{B+B}. That is the basis
for my strong belief that a similar reduction to hypergeometric series, or
some generalization~\cite{AKdF,HE,SK,BBBB} of them, underlies the finding
of Laporta and Remiddi that the three-loop anomaly involves just one Euler
sum with depth $k>1$, namely the very specific polylogarithmic
combination\Eq{U31}. It is also reasonable to expect that this reducibility
to Euler sums persists beyond three loops, where only numerical~\cite{TK1}
estimates are currently available.

\subsection{Knot theory}

To plain knot theorists, the preoccupations of a
knot/field-theorist~\cite{DK1,DK2,DK}
may appear rather restricted. The correspondence claimed
by Dirk Kreimer is between {\em positive\/} knots and
the numbers appearing in field-theory counterterms.
His process of discovery began
with the observation~\cite{DK1} that the removal of
sub-divergences from Feynman diagrams,
construed as a skein relation between link diagrams encoding
the momentum flow, yields a
counterterm that is rational if the skeining results
in the unknot,
as in the case of ladder diagrams, where
cancellations of $\zeta$-functions have been demonstrated
perturbatively~\cite{DK2} and non-perturbatively~\cite{DKT}.
It continued~\cite{DK2} with
the observation that the 2-braid torus knot $(2L-3,2)$ is
produced by skeining the crossed-ladder diagram that generates~\cite{BU}
$\zeta(2L-3)$ at $L\geq3$ loops.

To continue the correspondence,
it was clearly necessary to do two things: to
go to loop-numbers, $L$, higher than the
then current limit of $L=5$~\cite{phi}, which is how I became
involved~\cite{AI95},
and to enumerate positive knots, which
we\footnote{Table~5 was obtained in collaboration
with Dirk Kreimer, who will give further details in~\cite{DK}.}
have taught REDUCE~\cite{RED}
to do, up to 17 crossings, barring degeneracy of HOMFLY~\cite{VFRJ}
polynomials.  Here attention is largely restricted to knots with
up to 13 crossings, as in Table~5.
Knots with more crossings will figure in~\cite{DK}.

{\bf Table~5}:
Positive prime knots related to Euler sums,
via field-theory counterterms.
\[\begin{array}{|l|l|l|}\hline
\mbox{crossings}&\mbox{knots}&\mbox{numbers}\\[3pt]\hline
2a+1
&\so{2a+1}&\ze{2a+1}\\\hline
8
&\fk{}3{}3=8_{19}&\kk53\\\hline
9
&\mbox{none}&\mbox{none}\\\hline
10
&\fk{}5{}3=10_{124}&\kk73\\
&\fk{}333=10_{139}&?\\
&\fk2233=10_{152}&?\\\hline
11
&\eltft&\kkk353\\\hline
12
&\fk{}7{}3&\kk93\\
&\fk{}5{}5&\kk75-\frac{\pi^{12}}{2^5\tm10!}\\
&\fk{}353&?\\
&\fk{}335&?\\
&\fk2235&?\\
&\fk2334&?\\
&\fk3333&?\\\hline
13
&\sk{}32{}42&\kkk373\\                                    
&\sk2{}3223&\kkk535\\                                     
&\sk{}3{}{}52&?\\                                         
&\sk2{}33{}3&?\\                                          
&\sk{}3{}233&?\\                                          
&\so2\st2\so{}(\se{}\st3)^2&?\\                           
&(\st{}\so{}\se{}\st{})^3\so{}&?\\                        
&(\st{}\so{}\se{}\st{})^3\st{}&?\\                        
\hline\end{array}\]

Excluding the 2-braid torus knots that
correspond to depth-1 Euler sums, the tally of positive knots with
8 to 13 crossings is
apparent in the knot column of Table~5. It forms
the sequence 1, 0, 3, 1, 7, 8, \ldots, beginning at 8 crossings
with the 3-braid torus knot $8_{19}$.
By contrast, the corresponding sequence for {\em all\/} prime knots
with 8 to 13 crossings is given by M0851 of~\cite{SP,NJAS} as
21, 49, 165, 552, 2176, 9988, \ldots, whose richness reveals the
specificity of the preoccupations of knot/number/field theorists,
to whom, of course, the sparsity
of positive knots is a delight: it is broadly commensurate with the
slow growth in number of irreducible Euler sums, at corresponding levels.

There is no method, as yet, to assign numbers to knots,
other than by brute-force evaluation of counterterms from diagrams whose
skeinings produce the knots. The obstacle to high-precision evaluation
of such counterterms, beyond 6 loops, is apparent from the fact
that the knots $10_{139}$ and $10_{152}$ result, at 7 loops, from
diagrams whose evaluation, via Gegenbauer-polynomial
techniques~\cite{CKT}, entails 7-fold summations weighted by
the squares~\cite{BGN} of 6--$j$ symbols, in such a manner as to make the
evaluation-time of a truncation at $N$ increase at least as fast as $N^4$.
This is in stark contrast with the linear growth in Section~4.1,
whose computational
challenge hence pales into insignificance, compared with that
required for~\cite{AI95}.
\begin{center}
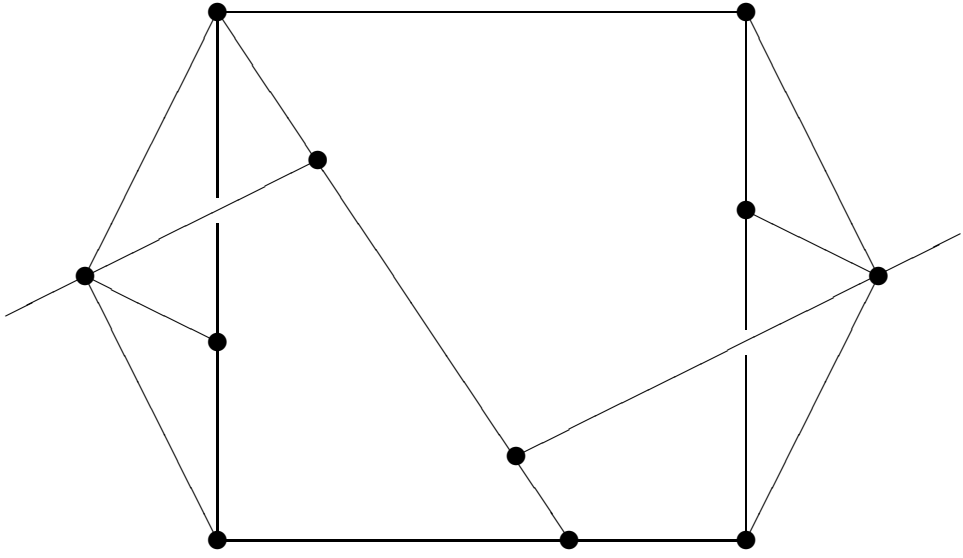
{\bf Fig.~1:} A diagram from cutting a 9-loop bubble that
skeins to 13-crossing 4-braids.\\
\begin{picture}(400,240)(-200,-125)
\put(-100, 100){\line( 1, 0){200}}
\put(-100,-100){\line( 1, 0){200}}
\put( 100,-100){\line( 0, 1){ 70}}
\put( 100, 100){\line( 0,-1){120}}
\put(-100,-100){\line( 0, 1){120}}
\put(-100, 100){\line( 0,-1){ 70}}
\put(-100, 100){\line( 2,-3){133}}
\put(-150,   0){\line( 1, 2){ 50}}
\put(-150,   0){\line( 1,-2){ 50}}
\put( 100, 100){\line( 1,-2){ 50}}
\put( 100,-100){\line( 1, 2){ 50}}
\put(- 62,  44){\line(-2,-1){118}}
\put(-150,   0){\line( 2,-1){ 50}}
\put( 150,   0){\line(-2, 1){ 50}}
\put(  13, -68){\line( 2, 1){168}}
\put( 100, 100){\vtx}
\put(-100,-100){\vtx}
\put( 100,-100){\vtx}
\put(-100, 100){\vtx}
\put(-150,   0){\vtx}
\put( 150,   0){\vtx}
\put(  33,-100){\vtx}
\put(-100, -25){\vtx}
\put(- 62,  44){\vtx}
\put(  13, -68){\vtx}
\put( 100,  25){\vtx}
\end{picture}
\end{center}

Fortunately, there is a class of Feynman diagrams that may, with effort,
be reduced to triple Euler sums, by analytical methods. A relatively
simple example is provided by
the 8-loop two-point diagram of Fig.~1, which is finite in 4 dimensions.
With unit external momentum, a massless propagator $1/p_{\rm line}^2$
for each line, unit vertices, and euclidean measure
$\pi^{-2}\int{\rm d}^4p_{\rm loop}$ for each loop,
this diagram was evaluated by REDUCE as
\begin{eqnarray}
G(3,2,2)&=&
-405\kkk373
+\df{3675}{4}\kkk292
+4680\kkk272
+\df{21285}{4}\ze{13}
-20535\ze{11}\nl
+6480\ze9\ze3
+12680\ze9
+480\ze7\ze5
-19500\ze7\zs3\nl
-7200\ze7\ze3
-79380\ze7
-14700\zs5\ze3
-1200\zs5\nl
-68160\ze5\zs3
+38880\ze5\ze3
-11520\zc3
+30240\zs3
\,,\label{G322}
\end{eqnarray}
by means of the master formula~\cite{AI95,5LB}
\begin{eqnarray}
G(a,b,c)&=&\sum_{i,j,k}{2a-i\choose a}{2b-j\choose b}{2c-k\choose c}
\frac{(i+j+k)!}{i!j!k!}\sum_{p,m,n}
\frac{\Delta(p,m,n)}{p^{2a-i}m^{2b-j}n^{2c-k}}\nonumber\\
&\times&\left[\left(\frac{2}{p+m+n-1}\right)^{i+j+k+1}+
\left(\frac{2}{p+m+n+1}\right)^{i+j+k+1}\right]\,,\label{gabc}
\end{eqnarray}
where $\Delta(p,m,n)$ results from angular integrations over
Chebyshev polynomials~\cite{CKT} and is 1, or 0, according
as whether $g=(p+m+n+1)/2$ is, or is not, an integer satisfying
$g>\max(p,m,n)$.

Each irreducible 3-fold sum appearing in\Eq{G322} belongs
to one of the two-parameter families:
\begin{eqnarray}
\kkk{2m+1}{2n+1}{2m+1}&=&
\zeta(2m+1,2n+1,2m+1)-\zeta(2m+1)\,\zeta(2m+1,2n+1)\nl
+\sum_{k=1}^{m-1}{2n+2k\choose2k}\zeta_P(2n+2k+1,2m-2k+1,2m+1)\nl
-\sum_{k=0}^{n-1}{2m+2k\choose2k}\zeta_P(2m+2k+1,2n-2k+1,2m+1)
\,,\label{K3o}\\
\kkk{2m}{2n+1}{2m}&=&
\zeta(2m,2n+1,2m)+\zeta(2m)\left\{\zeta(2m,2n+1)+\zeta(2m+2n+1)\right\}\nl
+\sum_{k=1}^{m-1} {2n+2k\choose2k  }\zeta_P(2n+2k+1,2m-2k,2m)\nl
+\sum_{k=0}^{n-1} {2m+2k\choose2k+1}\zeta_P(2m+2k+1,2n-2k,2m)
\,,\label{K3e}
\end{eqnarray}
with product terms
\begin{equation}
\zeta_P(a,b,c)=\zeta(a)\left\{2\,\zeta(b,c)+\zeta(b+c)\right\}
\,,\label{zp}
\end{equation}
whose systematic inclusion, with the combinatoric
factors in\Eqq{K3o}{K3e}, removes
all trace of the non-knot number $\pi^2$ from\Eq{G322}, and likewise
from {\em every} diagram $G(a,b,c)$ with
$a+b+c\leq11$, according to similar, but {\em much\/} lengthier, evaluations
of two-point functions obtained by cutting bubble diagrams with
up to 13 loops, corresponding to knots with up to 23 crossings.

In general, subdivergence-free bubble diagrams, with up to $L$ loops,
yield the knot-numbers $\{\kkk{a+2}{2b+1}{a+2}\mid a\geq0,\ b\geq0,\
L-4\geq a+b\}$ as the very specific combinations\Eqq{K3o}{K3e}
of Euler sums with depths $k\leq3$ and levels $l\leq2L-3$~\cite{AI95}.
They are knot-numbers, in the sense of~\cite{DK1,DK2,AI95},
because counterterms, from subdivergence-free
diagrams that skein to produce the corresponding knots,
contain these numbers, and products of other knot-numbers,
corresponding to factor~\cite{CCA} knots. The counterterms do {\em not}
contain the non-knot irreducibles $\ln2$ and $\pi^2$. Thus the combinations
of Euler sums in\Eqq{K3o}{K3e} provide log-free and $\pi$-free bases
for search spaces in which to evaluate classes of counterterms from
diagrams that skein to 4-braids. For example, it was possible~\cite{AI95}
to evaluate all 7-loop $\phi^4$ counterterms
from subdivergence-free diagrams that
skein to 4-braids in terms of just 3 knot-numbers: $\ze{11}$,
$\ze5\zs3$, and $\kkk353+7\ze5\zs3=\zeta(3,5,3)-\ze3\zd53$, corresponding
to the 2-braid torus knot $(11,2)$, the factor knot $5_1\times3_1\times3_1$,
and the uniquely positive 11-crossing
hyperbolic\footnote{In~\cite{AI95}, this knot, $10_{139}$, and $10_{152}$,
were wrongly called satellite knots; all three are, in fact, hyperbolic.}
knot $\eltft$. The factor knot $3_1\times8_{19}$ is not produced by
skeining the link diagrams that encode the momentum flow.
However, every diagram that skeins to $\eltft$ also produces
the other two knots, $(11,2)$ and $5_1\times3_1\times3_1$.
Until a method is devised
to predict the rational coefficients with which knot-numbers occur,
rather than determining them empirically, as at present, the association
$\eltft\simeq\kkk353$ can be made only modulo $\ze{11}$
and $\ze5\zs3$ terms. On this understanding, $\kkk353$ appears in the
11-crossing entry of Table~5.

In Table~5, $\kkk353$ is the {\em first\/} irreducible triple sum
to appear, at 11 crossings.
That is because the knot-numbers\Eqq{K3o}{K3e} are not
all independently irreducible. Up to level 13, the following
odd-zeta reductions are obtained, by extending the
methods of~\cite{BG} to\Eq{gabc}:
\begin{eqnarray}
\kkk212&=&\df{9}{2}\ze5\nq
\kkk232&=&\df{75}{8}\ze7\nq
\kkk313&=&-\df{1}{4}\ze7\nq
\kkk252&=&\df{439}{36}\ze9+\df{8}{3}\zc3\nq
\kkk333&=&\df{1}{3}\ze9-\df{4}{3}\zc3\nq
\kkk414&=&\df{115}{18}\ze9-\df{4}{3}\zc3\nq
4\kkk353-5\kkk272&=&-\df{1031}{24}\ze{11}-88\ze5\zs3\nq
\kkk434-5\kkk272&=&\df{103}{8}\ze{11}-80\ze5\zs3\nq
2\kkk515-5\kkk272&=&-\df{1091}{24}\ze{11}-56\ze5\zs3\nq
32\kkk454+140\kkk373-525\kkk292&=&
\df{24425}{3}\ze{13}-12880\ze7\zs3-8400\zs5\ze3\nq
64\kkk535-100\kkk373+175\kkk292&=&
-673\ze{13}+4400\ze7\zs3+3440\zs5\ze3\nq
8\kkk616-12\kkk373+49\kkk292&=&
-115\ze{13}+976\ze7\zs3+752\zs5\ze3\,.
\label{krel}
\end{eqnarray}
At level $l=2L-3>3$, corresponding to $L>3$ loops,
there are $L-3$ knot-numbers of the form
$\{\kkk{a+1}{2b+1}{a+1}\mid a>0,\ b\geq0,\ a+b=L-3\}$,
while\Eq{4b} gives $\lceil(L-3)^2/12\rceil-1$ irreducibles~\cite{BG}.
Hence the knot-numbers\Eqq{K3o}{K3e} fail to exhaust the irreducibles,
for $L\geq16$ loops. Up to $L=13$ loops, their sufficiency has been
proven, using REDUCE.

Two independent irreducibles, chosen to be
$\kkk373$ and $\kkk535$,
appear in the 13-crossing part of Table~5. They are associated with the
first two braid words, on the basis of intensive
skeining of the link diagrams that encode momentum flows in bubble diagrams.
A hint of the pen-and-paper labour undertaken by Dirk Kreimer is given by
his drawings in~\cite{BGK}, which refer to much simpler Feynman diagrams.
Since at least two 13-crossing knots of Table~5 emerge from
skeining any diagram that yields irreducible triple sums of level 13,
we cannot yet determine which rational
combination of the two irreducibles is associated with a given
knot.
By a combination of skeining and inspection of factorizations of
Alexander~\cite{VFRJ} polynomials, we arrive at the association
of the first two 13-crossing knots of Table~5 with, as yet undetermined,
linear combinations of $\kkk373$ and $\kkk535$.
The remaining 6 knot-numbers at level 13
cannot all come from $S_{13,3}$, since Table~1 reveals that $E(13,3)=7$,
of which the two non-alternating irreducibles have already been
accounted for.
It is possible that further 13-crossing knot-numbers come from
$S_{13,5}$, whose size, $S(13,5)=133$, puts it out of the
reach of MPPSLQ for the foreseeable future.

At even levels, the associations of Table~5 are made with combinations
\begin{equation}
\kk{a}{b}=\zeta(-a,b)-\zeta(-b,a)\,,\label{K2}
\end{equation}
of {\em alternating\/} double Euler sums. The very simple constructs
$\kk53$ and $\kk73$ remove $\{\pi^{2n}\mid n\leq5\}$ from
diagrams with up to 7 loops that skein to $8_{19}$ and $10_{124}$. Modulo
terms corresponding to factor knots, these two knot-numbers agree with the
findings of~\cite{AI95}, where results were written, equivalently, in terms
of $29\ze8-12\zd53$ and $94\zd73-793\ze{10}$. Previously, the numbers in
these combinations appeared gratuitous; now they are seen as consequences
of ignoring the wider world of alternating sums.

Calculations at
higher loops reveal that $\kk{2a+5}{3}$ is a knot-number at level $2a+8$.
However, a multiple of $\pi^{2(a+b+6)}$ must be subtracted from
$\kk{2a+7}{2b+5}$ to obtain a knot-number for 3-braids with 12 or more
crossings. Table~5 shows that this subtraction has a very simple form
at 12 crossings. The discovery of the general form of the subtraction
is frustrated by the fact that counterterms associated with even-crossing
knots are generally much more difficult to calculate than those
associated with odd numbers of crossings, as witnessed by
the table of results in~\cite{AI95}, where $10_{139}$ and $10_{152}$
conspired to frustrate the precise evaluation of all
subdivergence-free contributions to the 7-loop $\beta$\/-function
of $\phi^4$-theory. As these knots entail computation times
scaling as $N^4$ for truncation at $N$, they leave
us with a 7-loop result that is known to `only' 11 significant
figures, after $10^3$ CPUhours. By way of a hard-won, but relatively simple,
exact numerical result at 8 loops,
\begin{equation}
M(2,2,1,1)=2^{14}\tm3\kk93+2^4\tm3\tm5^3\tm7\ze3\ze9
-2^5\tm3^3\zf3-2^3\tm3^2\tm577\ze5\ze7\label{M2211}
\end{equation}
is offered, as the finding of MPPSLQ for the Feynman diagram of Fig.~2.
\begin{center}
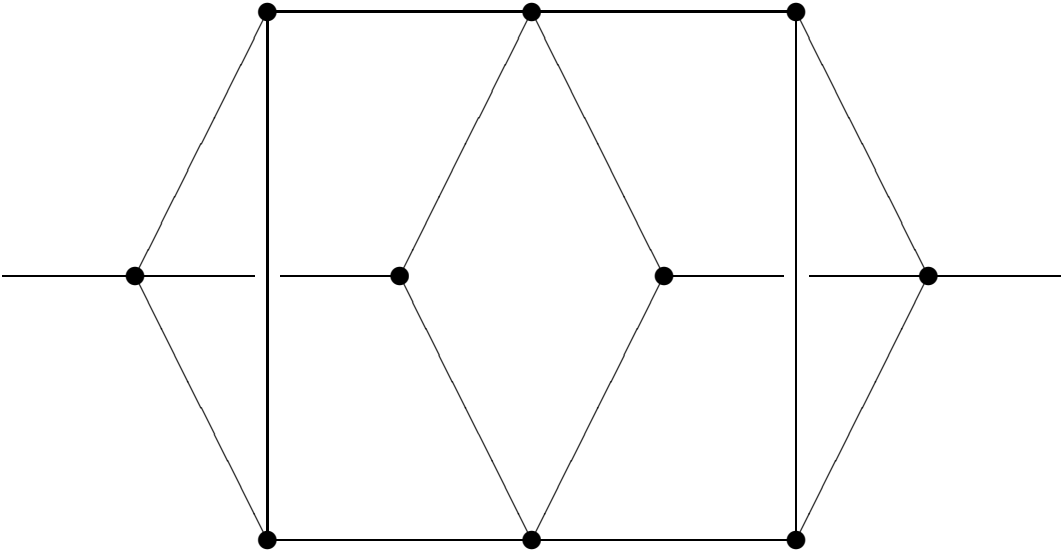
{\bf Fig.~2:} A diagram from cutting an 8-loop bubble that
skeins to 12-crossing 3-braids.
\begin{picture}(400,240)(-200,-125)
\put(-100,+100){\line(1, 0){200}}
\put(-100,-100){\line(1, 0){200}}
\put(+100,-100){\line(0, 1){200}}
\put(-100,-100){\line(0, 1){200}}
\put(-200,   0){\line(1, 0){ 95}}
\put(- 95,   0){\line(1, 0){ 45}}
\put(+ 50,   0){\line(1, 0){ 45}}
\put(+105,   0){\line(1, 0){ 95}}
\put(-150,   0){\line(1, 2){ 50}}
\put(- 50,   0){\line(1, 2){ 50}}
\put(-150,   0){\line(1,-2){ 50}}
\put(- 50,   0){\line(1,-2){ 50}}
\put(   0, 100){\line(1,-2){ 50}}
\put(   0,-100){\line(1, 2){ 50}}
\put( 100, 100){\line(1,-2){ 50}}
\put( 100,-100){\line(1, 2){ 50}}
\put(+100,+100){\vtx}
\put(-100,-100){\vtx}
\put(+100,-100){\vtx}
\put(-100,+100){\vtx}
\put(   0,+100){\vtx}
\put(   0,-100){\vtx}
\put( -50,   0){\vtx}
\put( +50,   0){\vtx}
\put(-150,   0){\vtx}
\put(+150,   0){\vtx}
\end{picture}
\end{center}

The evaluation\Eq{M2211} was accomplished by using 4-dimensional
Chebyshev-polynomial expansions, derived in~\cite{lad},
for the iteratively-defined coordinate-space constructs
\begin{equation}
P_{n+1}(x,y)=\int\frac{{\rm d}^4z}{\pi^2z^2}\,P_n(x,z)\,P_0(z,y)\,;
\quad P_0(x,y)=1/(x-y)^2\,,
\label{Pn}
\end{equation}
which are then combined by REDUCE to perform a radial integration in
\begin{equation}
M(a_1,a_2,a_3,a_4)=x^4\int\frac{{\rm d}^4y}{\pi^2}\prod_i P_{a_i}(x,y)
=\sum_{n_i}A\left(\{n_i\}\right)\,R\left(\{a_i\};\{n_i\}\right)\,,
\label{Mform}
\end{equation}
giving~\cite{5LB} a 4-fold sum, whose radial term, $R$, is a huge expression
involving inverse powers of $n_i$ and $h=\frac12\sum_i n_i$, while the
angular term, $A$, vanishes unless $h$ is an integer greater than any
$n_i$, in which case $A$ is the smallest of the 8 values $n_i$ and $h-n_i$.
Taking the REDUCE result, TRANSMP produces MPFUN code that implements the
brute-force accelerator\Eq{pos}, with demon-number $d=2$, yielding enough
significant figures for MPPSLQ to discover\Eq{M2211}. It would be very hard
to obtain such results analytically. As so often, in field theory, the
whole is much simpler than the parts: thousands of 4-fold non-Euler sums
produce a one-line, $\pi$-free result, in terms of the alternating
double-sum knot-number $N_{9,3}$, which cannot be expressed in terms of
non-alternating sums. This led me to seek and find such things as\Eq{zu93},
which in turn led to the enumeration\Eq{Elk}.

Clearly there is a pressing need for a more developed knot/number/field
theory, which might tell one which Euler (or other) sums in counterterms to
associate with which knots, without need of laborious calculations of
Feynman diagrams. In particular I would dearly like to know whether the two
undetermined knot-numbers at level 10 are irreducible alternating 4-fold
Euler sums, residing in $S_{10,4}$. Analytical assistance is ardently
sought; without it, additional numerical work is likely to add little
understanding.

Eventually, it may prove possible to relate Euler sums to positive knots,
directly, in a way that is consistent with the field-theory route, yet does
not oblige one to follow it. That is, I suggest, a substantial task, since
it is notoriously difficult to derive non-trivial statements that apply to
all members of a well-defined class of knot. For example, the enumeration,
by REDUCE, of positive knots up to 17 crossings, is insecure against the
possibility that two distinct positive knots might have the same HOMFLY
polynomial, though no example of positive mutation has come to light.

It may be that the calculational complexities of field-theory counterterms,
and the classificational complexities of knot theory, are mutual echoes,
with which the now-apparent combinatoric simplicity of the enumeration of
irreducible Euler sums eventually fails to resonate. For the present, however,
the following correspondences, from computations to 13 loops,
are rather impressive, to my mind.
\begin{enumerate}
\item Barring cancellations between diagrams, associated with dynamical
symmetries, such as local gauge invariance~\cite{BDK} or
supersymmetry~\cite{BGK},
the Euler sum $\zeta(2L-3)$, corresponding~\cite{DK2} to the 2-braid
torus knot $(2L-3,2)$, first appears in anomalous dimensions at $L$
loops.
No other irreducible results from subdivergence-free diagrams
with less than 6 loops, because no other positive knot has
less than 8 crossings.
\item At 6 loops, 3-braids start to appear. The first of these is
$8_{19}\simeq\kk53$. The irreducibility of its knot-number
was confirmed in~\cite{BBG}, with no knowledge
of prior developments in field theory~\cite{exp,zs6}.
\item At 7 loops, 4-braids start to appear. The first of these is
$\eltft\simeq\kkk353$. The irreducibility of its knot-number
was confirmed in~\cite{BG},
following communication of its appearance in field theory~\cite{AI95}.
The 3-braid $10_{124}\simeq\kk73$ is also encountered at 7 loops~\cite{AI95},
in accord with the tally of~\cite{BBG}.
\item At 8 loops, there appear:
a pair of 13-crossing 4-braids, with knot-numbers
$\kkk373$ and $\kkk535$, in {\em accord\/} with the tally of~\cite{BG};
and a pair of 12-crossing 3-braids, with knot-numbers
$\kk93$ and $\kk75\mbox{ modulo }\pi^{12}$,
in {\em excess\/} of the tally of {\em non\/}-alternating double
sums in~\cite{BBG}.
The latter pair led, via\Eq{M2211}, to\Eq{zu93}, which resolved the
apparent conflict between knot/field theory and number theory and
then produced the enumerations\Eqq{Elk}{El}, as
contributions from mathematical physics to pure mathematics.
\item Results up to 13 loops confirm the association of the
knot-numbers\Eqq{K3o}{K3e} with 4-braids, up to 23 crossing.
\item Results on 14-crossing knots, appearing at 9 loops,
will be given in detail in~\cite{DK}.
They confirm the appearance of the expected pair of
double-sum irreducibles,
$\kk{11}3$ and $\kk95\mbox{ modulo }\pi^{14}$. For the first time,
a {\em truly\/} irreducible 4-fold Euler sum is obtained from a
Feynman diagram.
The associated knot-number is
$\zeta(5,3,3,3)+\zeta(3,5,3,3)-\ze3\zn533\mbox{ modulo }\pi^{14}$.
\end{enumerate}

\subsection{Number theory}

Mathematicians have only recently, it appears, made significant
extensions of Euler's original study of double sums~\cite{OO}.
Alternating double sums, familiar in field theory~\cite{VAS}
since~\cite{exp},
were studied in~\cite{BBG}; triple sums, encountered
in~\cite{zs6}, were studied in~\cite{BG,CM}; generic
non-alternating sums were studied in~\cite{AG},
where the sum rule\Eq{AG} was obtained.

The extension of these studies into the entire domain of $k$-fold Euler sums,
at all levels $l$, with all possible alternations of sign,
was undertaken, in this work, in an unashamedly
experimental\footnote{See~\cite{BBGP} for a suggestion of a working
definition of {\em experimental\/} mathematics.}
manner, stemming from an urgent need further to develop the connection
between knot theory and quantum field theory.

Mathematics, at its purest, relies on little more than the fertile
invention of the human mind. Mathematical physics often spawns
structures of even greater beauty, thanks to what Einstein called
the `incomprehensible' comprehensibility of the natural world.
As a new variation on this oft-repeated theme, field theory has led,
from the observation~\cite{DK1} of the rationality of ladder-diagram
counterterms, and the skeining of zeta-rich crossed-ladder
diagrams~\cite{DK2}, to a uniquely positive hyperbolic 11-crossing knot
in 7-loop~\cite{AI95} counterterms, and in this work to
the Feynman diagram of Fig.~2, whose evaluation\Eq{M2211} then
resulted in the
purely mathematical discovery that non-alternating Euler sums
require alternating sums for their reduction, as witnessed by\Eq{zu93}.
Setting field theory aside, for a brief while, the larger universe of
alternating Euler sums proved much easier to enumerate than its
non-alternating restriction, as witnessed by\Eq{Elk}.

The validity of the enumeration\Eq{Elk} is provisional: a year of
work and $10^3$ CPUhours, of the most exhaustive tests of which I and the
engines at my disposal are capable, fail to reveal the slightest flaw
in it. I invite colleagues with larger numerical appetites to test it
further, in the lively expectation that it will survive.

What is needed now is the closest thing to proof for which it is
reasonable to hope: the establishment by deductive methods of
the validity of\Eq{Elk} as an {\em upper bound\/}
on the number of irreducible
$k$-fold Euler sums at level $l$. The more ambitious aim of proving it as
an identity is unrealistic, until someone develops the machinery for proving,
{\em inter alia\/}, the irrationality of $\zeta^2(137)/\zeta(274)$, and
a denumerable infinity of suchlike things. The more modest proposal of
proving that the number of irreducibles is {\em no more\/} than that given
by\Eq{Elk}
seems eminently realistic. To anyone disposed to undertake it, I offer
the following informal restatement:
\begin{enumerate}
\item For concrete values of $l$ and $k$, such that $l+k$ is even,
and $l\geq k\geq1$, form all the partitions of $l$ into
precisely $k$ odd integers.
\item For each partition $p_i$, count the distinguishable
permutations of these odd integers and denote the answer by $P_i$.
\item Let $A_i$ be the number of products of lower-level irreducibles,
associated with partition $p_i$ by the generator\Eq{gen}
operating on already established irreducibles.
\item Summing $E_i=P_i-A_i$ over the odd partitions one arrives at\Eq{Elk}.
\end{enumerate}
By way of example, consider $S_{10,4}$, which admits of the odd
partitions $p_1=7+1+1+1$, $p_2=3+3+3+1$, and $p_3=5+3+1+1$, with
$P_1=P_2=4!/3!=4$ and $P_3=4!/2!=12$. To exemplify the structure,
only the argument strings of Section~6 are notated, with $(\ovl1)$ standing
for $-\ln2$.
Thus the 3 products
\begin{equation}
(7)(\ovl1)(\ovl1)(\ovl1),\,
(\ovl7,\ovl1)(\ovl1)(\ovl1),\,
(7,\ovl1,\ovl1)(\ovl1),\label{p1}
\end{equation}
leave $E_1=4-3=1$ as the number of irreducibles associated with $p_1$.
Similarly,
\begin{equation}
(3)(3)(3)(\ovl1),\,(\ovl3,\ovl1)(3)(3),\,(3,\ovl1,\ovl3)(3),\label{p2}
\end{equation}
leave $E_2=4-3=1$ irreducibles. Finally, the 9 products
\begin{eqnarray}&&
(5)(3)(\ovl1)(\ovl1),\,
(\ovl5,\ovl3)(\ovl1)(\ovl1),\,
(\ovl5,\ovl1)(3)(\ovl1),\,
(\ovl3,\ovl1)(5)(\ovl1),\,
(\ovl3,\ovl1)(\ovl5,\ovl1),\nl
(5,\ovl1,\ovl1)(3),\,
(3,\ovl1,\ovl1)(5),\,
(5,\ovl1,\ovl3)(\ovl1),\,
(3,\ovl1,\ovl5)(\ovl1),\label{p3}
\end{eqnarray}
leave $E_3=12-9=3$ irreducibles. The tally $E(10,4)=1+1+3=5$ agrees
with\Eq{Elk}. Note that in\Eqqq{p1}{p2}{p3}
the specific choices of Section~6 were made,
for irreducibles in spaces of lesser depth and level.
However, that was not necessary: all one needs is the {\em number\/} of
irreducibles, associated with the sub-partition in the smaller space,
which is itself determined by the Aufbau.
By the same token, all one needs to carry forward from $S_{10,4}$,
for later stages of the Aufbau, are the numbers, $E_i$, of irreducibles
associated with each of the 3 partitions of 10 into 4 odd integers;
it is not necessary to specify a concrete basis by choosing signs
or specific orderings of arguments.
Note also that the total number of permutations
is merely an element of Pascal's triangle:
$\sum_i P_i = {\frac{l+k-2}{2}\choose k-1}$, in general, with
$4+4+12={6\choose3}$ in $S_{10,4}$.
It is, therefore, an elementary exercise to build up Euler's
triangle of Table~1, by repeated application of $E_i=P_i-A_i$, which
is, so to speak, the microscopic version of the
macroscopic results\Eqqq{Elk}{El}{Plk}. To paraphrase:
\begin{quotation}\noindent
The number of irreducibles is given, by
{\em Euler's\/} triangle, as the deficit between
the number that {\em Pascal's\/} triangle generates,
by permutation of integers in odd partitions, and
the number of products {\em already\/} given by previous deficits.
\end{quotation}
In its final form, a proof of\Eq{Elk}, as a rigorous upper bound,
may appear almost as simple as the
``Euler = Pascal -- Already'' paraphrase of what is to be proved.
It may, however, require formidable combinatoric
intuition to generate and, more importantly, to {\em organize\/}
sufficient permutation and partial-fraction
relations, between $k$-fold Euler sums, with all possible alternations of
sign, as effectively
as the authors of~\cite{BBG,BG} organized the relations between
{\em non\/}-alternating sums at depths $k=2$ and $k=3$. Mindful of
what was entailed by these lesser tasks,
I did not attempt the greater.

\subsection{Conclusion}

I marvel that the quantum electrodynamics~\cite{TK2}
of Dyson, Feynman, Schwinger, and Tomonaga~\cite{SSS} leads, after
50 years of dedicated calculation and the attainment of equally impressive
experimental accuracy~\cite{VSD},
to agreement~\cite{LR} between theory and experiment on the {\em eleventh\/}
significant figure in the magnetic moment of the electron.
That skillful calculator of double sums,
Leonhard Euler~\cite{LE}, would smile, one feels,
on seeing nothing more complicated than
$\sum_{n>m}(-1)^{n+m}/n^3m$
in the sixth-order perturbation expansion.
Gauss, too, might be amused to see numbers, from his hypergeometric
series~\cite{WP2,GHH}, and knots, whose codification~\cite{CCA} he began,
walk hand in hand, down the perturbation expansions of quantum field
theory, to all~\cite{BGK} orders in the coupling constant.

Such reassuring order, in the mathematical description of
nature, at astonishing levels of accuracy, reinforced by
experience~\cite{BGK,lnf,exp,zs6,WP2,BB,BFT,GG2} with hypergeometric series
generated by dimensional regularization, and now by the remarkable
matches between Euler sums~\cite{BBG,BG}
and Dirk Kreimer's knot/number/field theory~\cite{DK1,DK2,AI95,BDK,BGK},
leads me to believe that the $k\leq3$ sums
in\Eq{to7}, and hence in\Eq{C34}, more than suffice for the electron's
anomaly at eighth order.
Certainly, they suffice for the reduction of all Euler sums with
levels $l\leq7$.
At any level, one has only to consult\Eq{Elk} for the tally of
irreducible $k$-fold Euler sums,
be they construed as calculational obstacles or
mathematical friends.

\subsubsection*{Acknowledgements}

It is a pleasure to thank my field-theory collaborators,
Pavel Baikov,
David Barfoot,
Bob Delbourgo,
Jochem Fleischer,
John Gracey,
Wolfgang Grafe,
Norman Gray,
Andrey Grozin,
Slava Ilyin,
Andrei Kataev,
Dirk Kreimer,
Karl Schilcher,
Volodya Smirnov, and
Oleg Tarasov,
for their contributions, evident below, to the genesis of this work.
Generous communication and advice, from
David Bailey,
Jon Borwein,
David Bradley,
Richard Crandall,
Roland Girgensohn,
Andrew Granville,
Tony Hearn,
Tom Kinoshita,
Tolya Kotikov,
Simon Plouffe,
Eduardo de Rafael,
Ettore Remiddi,
Kurt Riesselmann, and
Neil Sloane,
advanced the project.
Travel grants from HUCAM (contract CHRX--CT94--0579) and PPARC
enabled consultation with colleagues in
Aspen, Barcelona, Bielefeld, Dubna, Karlsruhe, Mainz, Moscow, Munich,
and Pisa.
Equally vital were the cooperation of fellow users
of my department's Alpha cluster,
and the expertise of its manager, Chris Stoddart.
Most especially, I thank Dirk Kreimer for his
gentle encouragement and tenacious skeining, without
which I could not have accomplished the
enumeration of irreducible Euler sums.

\newpage
\raggedright

\end{document}